\documentclass[trackchanges]{aastex7}

\newcommand{\kms}{${\rm km\;s^{-1}}$}

\newcommand{\msun}{M$_\odot$}

\usepackage{graphicx}
\usepackage{color}
\usepackage{wrapfig}

\shorttitle{SQ-A: A Collision Triggered Starburst} 
\shortauthors{C.K. Xu et al.}

\newcommand{\lsim}{\; \lower2truept\hbox{${< \atop\hbox{\raise4truept\hbox{$\sim$}}}$}\;}
\newcommand{\gsim}{\; \lower2truept\hbox{${> \atop\hbox{\raise4truept\hbox{$\sim$}}}$}\;}

\begin{document}


\title{SQ-A: A Collision Triggered Starburst in Intra-Group Medium of Stephan's Quintet}

\correspondingauthor{Cong Kevin Xu}

\author[0000-0002-1588-6700]{C.\,K. Xu}
\affiliation{Chinese Academy of Sciences South America Center for Astronomy, National Astronomical Observatories, CAS, Beijing 100101, China}
\affiliation{National Astronomical Observatories, Chinese Academy of Sciences, 20A Datun Road, Chaoyang District, Beijing 100101, China}
\email{xucong@nao.cas.cn}

\author[0000-0003-0202-0534]{C. Cheng}
\affiliation{Chinese Academy of Sciences South America Center for Astronomy, National Astronomical Observatories, CAS, Beijing 100101, China}
\affiliation{National Astronomical Observatories, Chinese Academy of Sciences, 20A Datun Road, Chaoyang District, Beijing 100101, China}
\affiliation{CAS Key Laboratory of Optical Astronomy, National Astronomical Observatories, Chinese Academy of Sciences, Beijing 100101, China}
\email{chengcheng@nao.cas.cn}

\author[0000-0001-7095-7543]{M.\,S. Yun}
\affiliation{Department of Astronomy, University of Massachusetts, Amherst, MA 01003, USA}
\email{myun@astro.umass.edu}

\author[0000-0002-7607-8766]{P.\,N. Appleton}
\affiliation{Caltech/IPAC, MC 314-6, 1200 E. California Blvd., Pasadena, CA 91125, USA.}
\email{apple@ipac.caltech.edu}

\author[0000-0003-2983-815X]{B.\,H.\,C. Emonts}
\affiliation{National Radio Astronomy Observatory, 520 Edgemont Road, Charlottesville, VA 22903, USA}
 \email{bemonts@nrao.edu}

\author[0000-0003-1740-1284]{J. Braine}
\affiliation{Observatoire de Bordeaux, UMR 5804, CNRS/INSU, BP 89, 33270 Floirac, France}
\email{jonathan.braine@u-bordeaux.fr}

\author[0000-0001-6217-8101]{S.\,C. Gallagher}
\affiliation{Institute for Earth and Space Exploration, Western University, 1151 Richmond St., London, ON N6A 3K7, Canada}
\email{sgalla4@uwo.ca}

\author[0000-0002-2421-1350]{P. Guillard}
\affiliation{Sorbonne Universit\'{e}, CNRS, UMR 7095, Institut d'Astrophysique de Paris, 98bis bd Arago, 75014 Paris, France}
\affiliation{Institut Universitaire de France, Minist\`{e}re de l'Enseignement Sup\'{e}rieur et de la Recherche, 1 rue Descartes, 75231 Paris Cedex 05, France}
\email{guillard@iap.fr}

\author[0000-0002-9471-5423]{U. Lisenfeld}
\affiliation{Departamento de F\'{i}sica Te\'{o}rica y del Cosmos, Universidad de Granada, 18071 Granada, Spain}
\affiliation{Instituto Carlos I de F\'{i}sica Te\'{o}rica y Computacional, Facultad de Ciencias, 18071 Granada, Spain}
\email{ute@ugr.es}

\author[0000-0002-5671-6900]{E. O'Sullivan}
\affiliation{Center for Astrophysics $|$ Harvard $\&$ Smithsonian, 60 Garden Street, Cambridge, MA, 02138, USA}
\email{eosullivan@cfa.harvard.edu}

\author[0000-0001-5073-2267]{F. Renaud}
\affiliation{
Observatoire Astronomique de Strasbourg, Universit\'e de Strasbourg, CNRS UMR 7550, F-67000 Strasbourg, France}
\affiliation{University of Strasbourg Institute for Advanced Study, 5 all\'ee du G\'en\'eral Rouvillois, F-67083 Strasbourg, France}
\email{florent.renaud@astro.unistra.fr}

\author[0009-0001-2178-4022]{P. Aromal}
\affiliation{Institute for Earth and Space Exploration, Western University, 1151 Richmond St., London, ON N6A 3K7, Canada}
\affiliation{Physics and Astronomy department, University of Western Ontario, 1151 Richmond Street, London, N6A 3K7, Ontario, Canada}
\email{apathaya@uwo.ca}

\author[0000-0003-3343-6284]{P.-\,A. Duc}
\affiliation{Universit\'{e} de Strasbourg, CNRS, Observatoire astronomique de Strasbourg (ObAS), UMR 7550, 67000 Strasbourg, France}
\email{pierre-alain.duc@astro.unistra.fr}

\author[0000-0002-0690-8824]{A. Labiano}
\affiliation{Telespazio UK S.L. for the European Space Agency (ESA), ESAC, Spain.}
\email{Alvaro.LabianoOrtega@ext.esa.int}

\author[0000-0001-5042-3421]{A. Togi}
\affiliation{Texas State University, 601 University Dr, San Marcos, TX 78666, USA}
\email{togiaditya@gmail.com}

\begin{abstract}
We present new observational evidence supporting the hypothesis that 
  SQ-A, a starburst in the intragroup medium (IGrM) of Stephan's Quintet (SQ),
  is triggered by a high-speed collision between two gas systems, one associated with the IGrM  ($\rm v \sim 6900$~\kms) and another with the intruder galaxy  NGC~7318b ($\rm v \sim 6000$~\kms). The new ALMA CO(2-1) dataset has angular resolutions  between  $0.2''$  and $7.0''$ and the new VLA HI datacube has an angular
  resolution of $\rm 6.6'' \times 7.9''$. The CO maps show that the two gas  systems are bridged by another system with an intermediate  velocity of $\sim 6600$~\kms, whereas the HI data show that  the component of $\rm v\sim 6600$~\kms  fits well into a  gap in the more extended $\rm v\sim 6000$~\kms component, albeit with a  displacement of $\sim 5$~kpc.  Both the bridge and the complementary  distributions between different gas systems are common features of  starbursts triggered by cloud-cloud collision. An analysis of clumps  (sizes of 100 -- 200~pc) reveals very diversified star formation (SF)  activity in clumps belonging to different kinematic systems, with the molecular gas depletion time of the $\rm v\sim 6900$~\kms clumps more than 10 times longer than  that of the $\rm v\sim 6600$~\kms clumps.  The results are  consistent with a scenario in which the enhanced SF activity (and  the starburst) in the system of $\rm v\sim 6600$~\kms is due to gas  compression generated in cloud-cloud collisions, whereas the  suppression of SF in the $\rm v\sim 6900$~\kms system is due to  vortices (i.e. gas rotation) generated in more complex collisions  involving dense clouds and diffuse intercloud gas accompanied by
  blast-wave shocks.  
\end{abstract}


\section{Introduction}\label{sec:intro}

Starbursts are rarely found in compact groups of galaxies which
  are aggregates of four to eight galaxies with implied space
  densities as high as those in cluster cores \citep{Hickson1982,
    Johnson2007}. This is remarkably different from galaxy mergers where
  interaction-induced starbursts are very common \citep{Sanders1996,
    Kennicutt1998a}. Indeed many interacting galaxies in compact
  groups have their star formation (hereafter SF) activity quenched
  rather than enhanced \citep{Alatalo2014}. SQ-A, a starburst in the
  intra-group medium (IGrM) of the famous compact group Stephan's
  Quintet (hereafter SQ), is quite unique.  As a prototype of compact
  groups, SQ is undergoing a very complex web of interactions between
  member galaxies and various constituents of the IGrM
  \citep{Moles1997, Sulentic2001}, which in turn have triggered some
  spectacular activities such as a $\sim$40~kpc large scale shock (shown as
  a large radio and X-ray ridge; \citealt{Allen1972, Xu2003,
    Trinchieri2003, OSullivan2009}) and the IGrM starburst SQ-A.  

SQ-A was discovered as a bright infrared source by
\citet{Xu1999} using the Infrared Space Observatory (ISO;
\citealt{Kessler1996}). Because of the relatively low resolution
($\sim 10''$) of the ISO observation, SQ-A originally included the
entire region beyond the northern tip of the large scale shock
\citep{Allen1972, Xu2003, Trinchieri2003, OSullivan2009}.
The SQ-A region, also known as the ``northern star burst region''
in the literature \citep{Gallagher2001, Fedotov2011}, is faint in optical and
near-infrared (NIR) continuum \citep{Xu1999} though relatively bright in H$_\alpha$
\citep{Arp1973, Moles1998, Iglesias-Paramo2001, MendesDeOliveira2001,
  DuartePuertas2019}.  In contrast it is very bright in the ISO 15$\mu m$ 
band, with a peak surface brightness second only to that of the Sy2 nucleus
  of the NGC 7319 which is the most massive galaxy of SQ.
The more recent observations, particularly
the optical spectroscopic imaging observations by
\citet{DuartePuertas2019}, reveal that the SQ-A region is dominated by a
compact starburst of size $\sim 5''$ ($\sim 2$ ~kpc) and a velocity $\sim 6680$ \kms
(hereafter SQ-A starburst), at the same time it contains many other
star-formation regions of different velocities.
According to \citet{Xu2005},
the total SFR of SQ-A is 1.34 $\rm M_\sun\; yr^{-1}$, comparable to that of
  the large Sb galaxy NGC~7319 (1.98 $\rm M_\sun\; yr^{-1}$). The very low NIR luminosity
  of SQ-A  indicates a young age ($\sim 10^7$~yr) for the starburst \citep{Xu1999}, consistent with the ages  of the star clusters found in the region \citep{Gallagher2001, Fedotov2011}.
Although the general morphology of the SQ-A region has been reproduced by numerical simulations \citep{Renaud2010, Hwang2012}, its starburst nature has not. 

There has been a debate on how the SQ-A starburst is triggered.
\citet{Xu1999} argued that the starburst is triggered by a collision
of two gas systems, one with system velocity of $\sim 6600$ \kms
associated with the IGrM and the other with velocity of $\sim
6000$ \kms associated with the intruder galaxy NGC~7318b, because
both the HI data \citep{Shostak1984} and $\rm H_\alpha$ data
\citep{Xu1999, MendesDeOliveira2001} show overlaps of the two systems
at the position of SQ-A. However, from the beginning, this
interpretation has been challenged by a difficult question: if 
high speed collision between gas systems can trigger starburst, why is
no starburst found in the shock region associated with
the large radio and X-ray ridge?
It is well established that the large scale shock is triggered by an ongoing collision
between the IGrM and the intruder NGC~7318b
\citep{Allen1972, Moles1997, Sulentic2001, Xu2003, Appleton2006, Guillard2009,
Cluver2010, Appleton2013, Guillard2022, Appleton2023}.
According to \citet{Cluver2010}, the shocks and
turbulence produced by the high-speed collision are more likely to
suppress the SF rather than trigger a
starburst. An alternative model to explain the starburst in SQ-A was proposed
by  \citet{MendesDeOliveira2001} who
listed the two kinematic systems in SQ-A as two separate tidal dwarf candidates,
suggesting that the SF in SQ-A is tidally induced and the collocation of
 multiple kinematic components in the region is due to projection effect.

Can shocks and turbulence powered by high-speed collisions
trigger starbursts and at the same time
suppress the SF activity elsewhere in the involved gas
systems? What are the physical mechanisms behind these processes?
The answer of these questions may have far-reaching inferences to our understanding
of the SF in gas systems, particularly the triggered SF \citep{Elmegreen1998}
which may play an important role in the SF activity in interacting galaxy systems.
In this paper, using the new high resolution CO data obtained by ALMA
and HI data by B, C \& D arrays of Karl G. Jansky Very Large Array (VLA),
we aim to answer these questions. In particular, we 
re-visit the question: is SQ-A triggered by a collision of the IGrM and the intruder? 
Significant progress has been achieved recently in cloud-cloud
collision (hereafter CCC) theories \citep{Habe1992, Takahira2014, Abe2021, Maeda2021,Maeda2024} which can provide powerful tools for our investigation. Encouraged by successful applications of the CCC models to large extra-galactic SF regions such as the overlap region of Antennae Galaxies and 30 Doradus in the
Large Magellanic Cloud \citep{Tsuge2021a, Tsuge2021b,Fukui2017,Tsuge2024}, we are motivated to find more signatures of collision triggered SF in SQ-A.

The paper is organized as follows: after this introduction, we present
the observational datasets used in this work in Section 2.  Our results
revealing new evidence for collisions between different gas systems in
SQ-A are presented in Section 3.  In Section 4 (and Appendix A) we
carry out an analysis on the properties of clumps (size $\sim 100$ -- 200~pc)
aiming to probe the cloud-level physical conditions leading to active
SF or suppression of it in SQ-A. 
 Section 5 is devoted to discussions.
In Section 6 we present our conclusions. 
Throughout the paper we assume a distance
of 94 Mpc (1'' corresponds to 0.45 kpc).
The velocities are in the optical redshift convention and
the local standard of rest (LSR) reference frame.

\section{Observations}\label{sec:Obs}

\subsection{ALMA Observations of CO(2-1)}\label{sec:ALMA}
CO(2-1) data of different angular resolutions and different coverages
are utilized in this paper.  They are obtained using ALMA in different
configurations. Mosaic observations using the Atacama Compact Array (ACA) cover a region of
$\sim 10$ arcmin$^2$ that includes all five member galaxies of SQ and most of
the IGrM, with a synthetic beam of $8.0''\times 7.0''$
\citep[see][]{Emonts2025}. We added to this new observations of similar mosaic coverage but higher ($\sim$0.6'') resolution done with the ALMA 12m array in C43-3 configuration from Aug 16 - Sept 5 2024 (ID: 2023.1.00177.S; PI: Appleton) with a total on-source integration time of 6.4 hours (hereafter TM2 data). These data were calibrated by the ALMA calibration pipeline version 2023.1.0.124 for Cycle-10 \citep{Hunter23} and delivered as calibrated MeasurementSets by the North American ALMA Science Center (NAASC). We used the Common Astronomy Software Applications (CASA) version 6.6.4.34 \citep{Casa22} to image the combined pointings of the ALMA and ACA mosaics that include the SQ-A region (TM2+ACA data). The imaging was done by using a Briggs weighting with robustness 0.5 \citep{Briggs95} and applying a multi-scale clean with scales 0, 5, 10, 25, and 60 pixels of 0.1'' down to a threshold of 3.5 mJy\,beam$^{-1}$. This resulted in a line-data cube with an angular resolution of $1.0''\times 0.66''$ at position angle (PA) -12.7$^{\circ}$, channel width of 15 \kms, and a root-mean-square (rms) noise of 0.85 mJy~beam$^{-1}$ in the center of the SQ-A field. A primary beam correction was applied, which increased the rms noise towards the edges of the field. 

Finally, the ALMA data of even higher
resolution in a single-pointing field centered
on the SQ-A starburst, namely the ``Field 4'' of \citet{Appleton2023},
were taken from the ALMA archive (ID: 2015.1.00241.S; PI: P. Guillard) and reprocessed with CASA using the default script {\sc ScriptForPI.py} to obtain the calibrated MeasurementSets. Then we clean the image by {\sc TCLEAN} with the parameters {\sc WEIGHTING = ``natural''}. The final data cube reaches an rms of about 0.66 mJy~beam$^{-1}$ and channel width of 4 $\rm km\,s^{-1}$, with a beam size of $0.38''\times 0.23''$ and  PA=-0.86 degree. Compared to the original Field 4 datacube presented in \citet{Appleton2023}, the reprocessed cube has the following improvements: (1) a velocity bin-width of 4 \kms instead of 20 \kms; (2) a much larger field of view of size = $38''$ (though results outside the $22''$ primary beam must be taken with caution). In this work, we assume a CO conversion factor $\rm \alpha_{CO10} = 1.0\; M_\sun (K\; km\; s^{-1} pc^{2})^{-1}$ and a CO(2-1) to CO(1-0)
luminosity ratio $\rm r_{21} = L'_{CO(2-1)} / L'_{CO (1-0)} = 0.34$
\citep{Emonts2025}. It is worth noting that both $\rm \alpha_{CO10}$ and $\rm r_{21}$ are quite uncertain. A detailed discussion of these uncertainties is given in Section~\ref{sec:conversion}.

\begin{figure*}[!htb]
\centering
\includegraphics[width=0.8\textwidth]{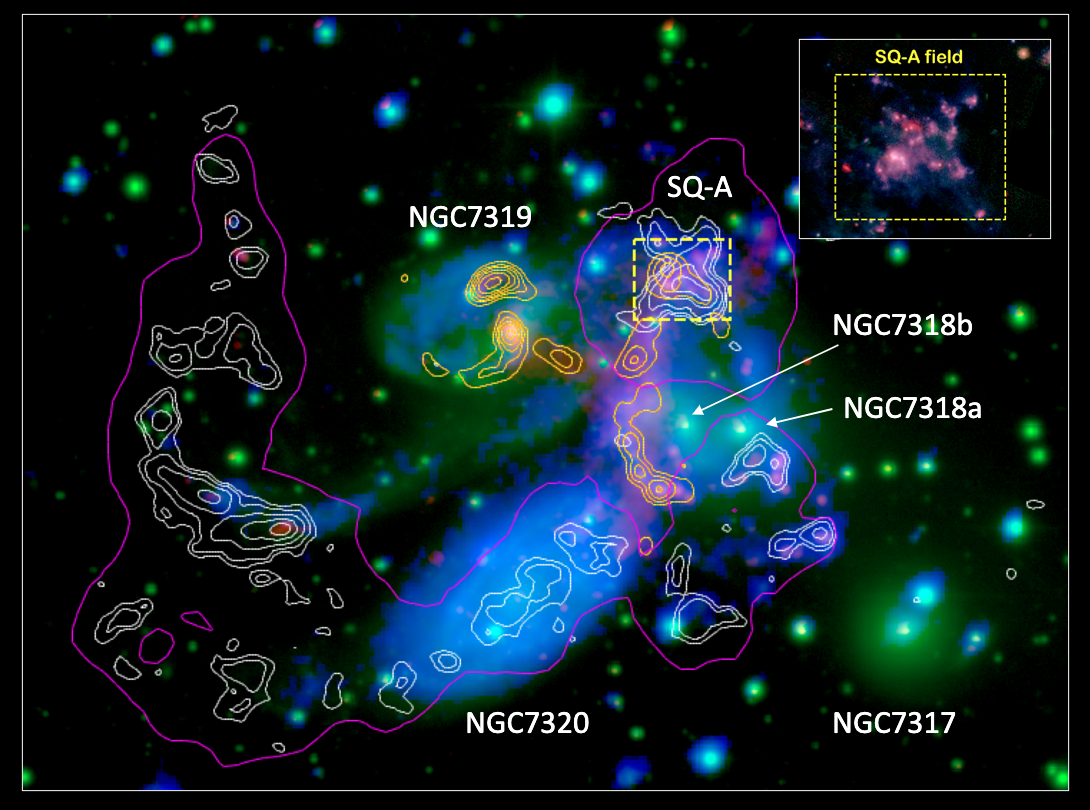}
\caption{
  Contour maps of the HI 21cm emission obtained using the VLA B, C \& D arrays (white contours, resolution: $7.9''\times 6.7''$) and the VLA C \& D arrays (magenta contours, resolution: $19.4''\times 18.6''$), and that of the CO (2-1) emission obtained using the ALMA ACA array (yellow contours, resolution: $8.0''\times 7.0''$) overlaid on a three-color image of the UV and optical emissions (blue: NUV, \citealt{Xu2005}; green: deep g-band, \citealt{Duc2018}; red: $\rm H_\alpha$, \citealt{Xu1999}). Note that the ACA CO (2-1) contours in the shock region look orange because of the colored background. The small box delineated by yellow dashed lines (size=$36''\times 30''$) marks the location of
the "SQ-A region"  studied in this paper. The inset in the upper-right corner is a three-color JWST image (blue: 7.7 $\mu m$; green: 10 $\mu m$; red: 15 $\mu m$) of the SQ-A region.}
\label{fig:fig1}
\end{figure*}

\subsection{VLA Observations of HI 21~cm Line}\label{sec:HI}
The VLA B-array observations of the 21 cm HI transition were
carried out for a total of 40 hours between May 27 and June 5 of 2005 as part of the AY155 program. The spectrometer setup used was identical to the earlier C \& D arrays observations reported by 
\citet{VerdesMontnegro2001} and  \citet{Williams2002}, with a velocity resolution of 21.5 km sec$^{-1}$.  A nearby VLA calibrator 2234+282 was observed frequently to track the instrument gain, and 3C~48 (0134+329) was observed for the bandpass and flux calibration. All of the data were calibrated using the standard VLA calibration procedure in AIPS and imaged using IMAGR.  In order to achieve better
UV coverage and better sensitivity, we combine these data with those
obtained with the VLA C \& D arrays (beam $=19.4''\times 18.6''$;
\citealt{Williams2002}).  The final HI data cube has a synthesized beam
of $7.9''\times 6.7''$ (FWHM) and a velocity coverage of 5550 -- 6885 \kms.

\subsection{JWST MIRI images}\label{sec:JWST}
Archived images in three MIRI bands of the James Webb Space Telescope (JWST)
are downloaded from the Dawn JWST Archive (DJA) website\footnote{\url{https://s3.amazonaws.com/grizli-v2/JwstMosaics/v4/index.html}}
\footnote{\url{https://zenodo.org/records/8370018}}
\citep{2023ApJ...947...20V}. These are F770W, F1000W and F1500W bands centered at 7.7$\mu m$, 10$\mu m$ and 15 $\mu m$ with the angular resolutions of $0.27'', 0.33''$ and $0.49''$, respectively \citep{Dicken2024}. The astrometric precision of the MIRI images matches well with that of GAIA DR3 which is on the order of 1 mas \citep{Libralato2024}.
 The F770W and F1000W data are used only in the inset of Fig.~\ref{fig:fig1}
        for illustration. The 15 $\mu m$ image taken from the F1500W data
is used in the analysis of clumps (Section~\ref{sec:clumps}). 
The 15 $\mu m$ emission is dominated by thermal emission of
warm dust heated by massive stars, and therefore is a good SF indicator \citep{Xu1999}.
In principle,  the 7.7$\mu m$ image
(dominated by the PAH emission) and 10$\mu m$ image (warm dust emission)
may also be used to study SF. They have better angular resolutions 
than the 15 $\mu m$ image and
are also insensitive to dust extinction which can be substantial in SQ-A
\citep{DuartePuertas2021}. However both F770W and F1000W are
significantly contaminated by shock-excited warm H$_2$ line emissions and therefore
are disfavored: the F770W filter includes the H2S(4) line at 8.025$\mu m$
and the F1000W filter the H2S(3) line at 9.665$\mu m$.
Both H$_2$ lines are bright in SQ \citep{Cluver2010, Appleton2023}.
It is worth noting that the F1500W filter also includes a bright emission line
$\rm [NIII]\lambda 15.56\mu m$ which is dominantly associated with SF regions. It
may not affect the 15 $\mu m$ flux as an SFR indicator significantly if
the $\rm [NIII]\lambda 15.56\mu m$-to-continuum ratio is relatively constant.

\section{Collision of Systems Associated with IGrM and Intruder}\label{sec:results}
In Fig.~\ref{fig:fig1} contours of the VLA HI maps of
resolution=$7.9''\times 6.7''$ (white contours) and of
resolution=$19.4''\times 18.6''$ (magenta contours), and contours of
the ACA CO (2-1) map (yellow contours, resolution = $8.0''\times 7.0''$)
are overlaid on a three-color image of the UV and optical emissions
(blue: NUV, \citealt{Xu2005}; green: deep g-band, \citealt{Duc2018};
red: $\rm H_\alpha$, \citealt{Xu1999}). A more comprehensive
comparison of the CO and HI data in entire SQ will be presented in a
future paper (Cheng et al., in preparation). In this paper we will
concentrate on the SQ-A region, shown by the small yellow box
(size=$36''\times 30''$) in the upper-right quadrant of the figure.
The inset in the upper-right corner is a three-color JWST image (blue:
7.7 $\mu m$; green: 10 $\mu m$; red: 15 $\mu m$;
\citealt{Pontoppidan2022}) of the SQ-A field which reveals the detailed morphology of the SF regions behind the CO and HI contours in the main figure.

\begin{figure}[!htb]
\centering
\includegraphics[width=0.4\textwidth]{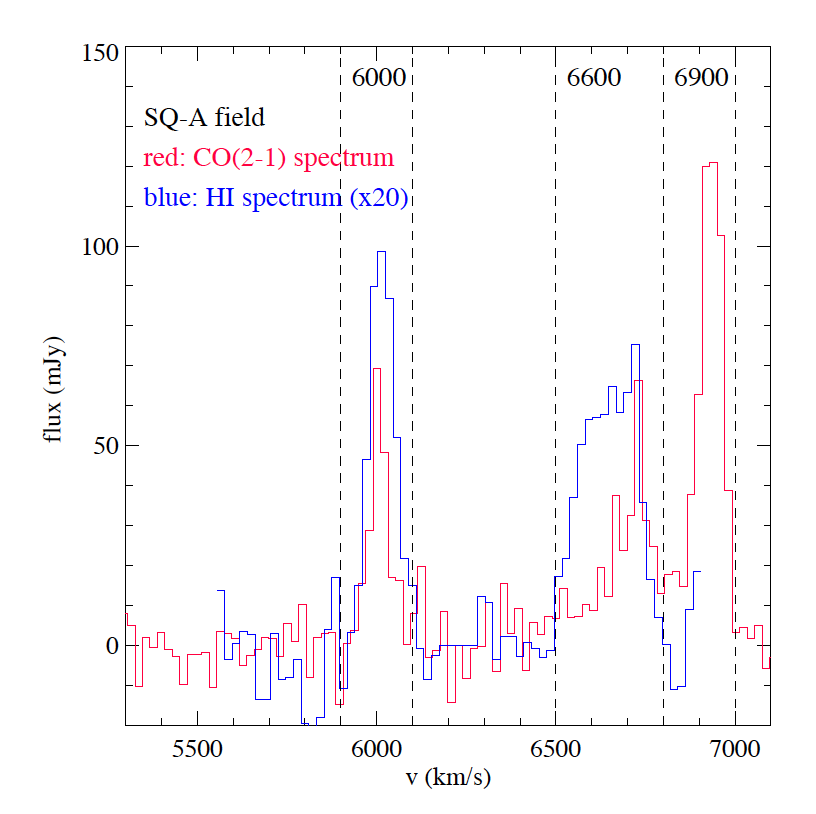}
\caption{
  Spectra of the CO (2-1) and the HI (scaled up by a factor of 20) in the SQ-A region.
  }
\label{fig:spectra}
\end{figure}

In Fig.~\ref{fig:spectra}, the CO(2-1) and HI spectra extracted from the CO(2-1) cube of ALMA ACA and the HI cube of VLA B, C \& D arrays in the SQ-A region (the yellow box in Fig.~\ref{fig:fig1})  are presented. It is worth noting that the two cubes have very similar angular resolutions (ACA CO(2-1): $8.0''\times 7.0''$, VLA B, C \& D arrays HI: $7.9''\times 6.7''$) and therefore the comparison between them is robust. The CO spectrum shows clearly three components.  Following the literature \citep{Shostak1984, Williams2002, Cheng2023}, we call them
the 6000, 6600 and 6900 components, which are in velocity ranges of 5900 -- 6100, 6500 -- 6800 and 6800 -- 7000 \kms, respectively. As shown in Table~\ref{tab:cold_gas}, the 6900 component has the highest molecular gas mass $\rm M_{mol}$ which is 1.9 and 1.5 times the $\rm M_{mol}$  of the 6000 and 6600 component, respectively.  For the 6000 and 6600 components the HI is much more abundant than the molecular gas: $\rm M_{HI}$ is about 10 times higher than $\rm M_{mol}$ for both components. Unfortunately the 
6900 component is not covered by the VLA HI observations. Preliminary results of a new MeerKAT observation of HI emission  in SQ (unpublished) indicate that the HI gas in SQ-A region does have a 6900 component, albeit with an $\rm M_{HI}$ less than half of that of the 6000 component (private communication with K.~Rajpurohit). Together with the data in Table~\ref{tab:cold_gas}, this suggests that the gas associated with the 6900 component is denser than that associated with the 6000 component because the former has a $\rm M_{mol}$-to-$\rm M_{HI}$ ratio $\sim 4$ times higher than the latter.   The HI 6900 component seems to be more extended than the CO counterpart but, given the larger beam of the MeerKAT map ($\sim 15''$) compared to that of the ACA, caution is needed about any direct comparison at this stage.

\begin{figure*}[!htb]
\centering
\includegraphics[width=0.4\textwidth]{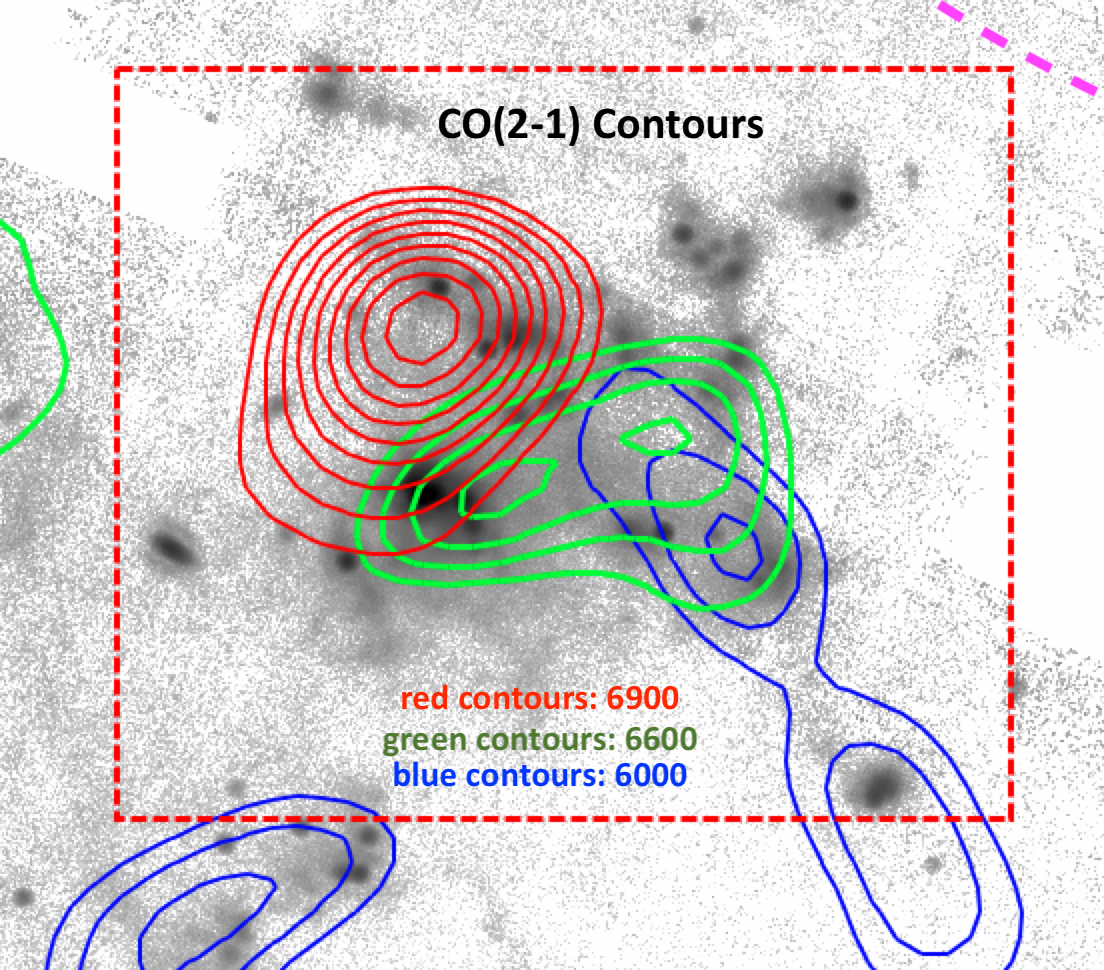}
\includegraphics[width=0.4\textwidth]{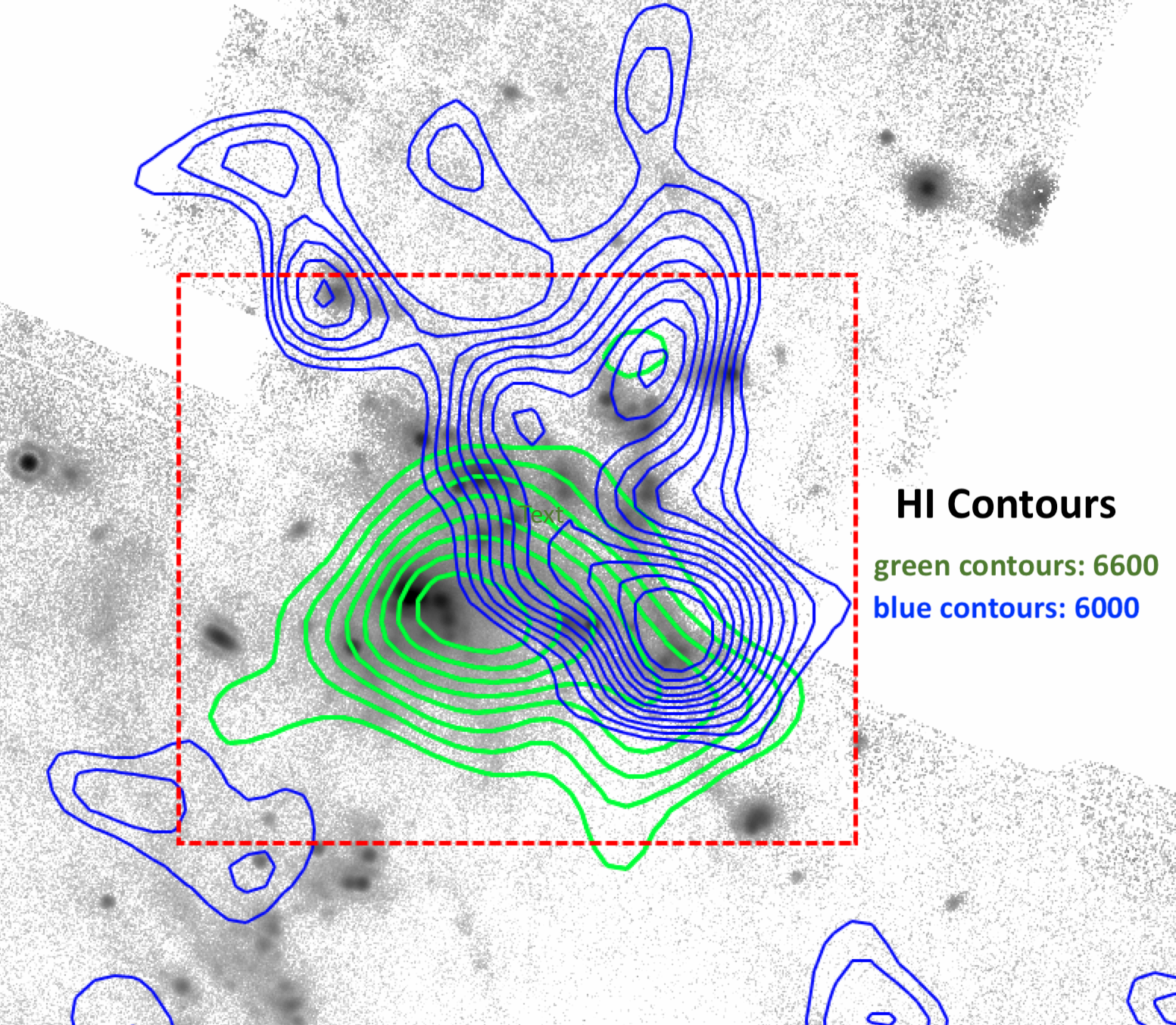}
\caption{
  {\it Left}:
  Contours of the ALMA ACA map of CO(2-1) in the SQ-A region, overlaid on a JWST 15 $\mu m$ image. The blue, green, and red contours are for the 6000, 6600, and 6900 components, respectively (contour levels: 1.3, 1.4, 1.5, … Jy~beam$^{-1}$~\kms).
  The three maps have r.m.s. of 0.36, 0.41, 0.39  Jy~beam$^{-1}$~\kms, respectively.
  The purple dashed line near the upper-right corner of the figure shows the boundary  of the ACA observation.
  The inner box delineated by the red dashed lines marks the locaton of the SQ-A region.
  {\it Right}:
  Contours of the VLA B, C \& D arrays map of HI overlaid on the JWST 15 $\mu m$ image. Blue and green contours are for the 6000 and 6600 components, respectively
  (contour levels at 0.04, 0.05, 0.06, …
 Jy~beam$^{-1}$~\kms). Both maps have the same r.m.s. = 0.01 Jy~beam$^{-1}$~\kms.  Again the red box delineates the SQ-A region.  }
\label{fig:ACA_HI}
\end{figure*}

\begin{table}
  \caption{Properties of three kinematic components of gas in SQ-A region. }
        {\centering
\resizebox{0.8\textwidth}{!}{%
\begin{tabular}{cccc}
\noalign{\smallskip} \hline \noalign{\medskip}
 & {\bf 6000}  & {\bf 6600} & {\bf 6900} \\
 \noalign{\smallskip} \hline \noalign{\medskip}
 v range [\kms] : & 5900 -- 6100    & 6500 -- 6800 &  6800 -- 7000 \\
 $\rm f_{CO(2-1)}$ [Jy~km~s$^{-1}]^{(a)}$ : &  4.8$\pm 0.5$ & 6.0$\pm 0.6$ & 8.8 $\pm 0.9$
 \\
 $\rm  M_{mol}\; [10^8 \times M_\sun]^{(b)}$ :      &      0.75$\pm 0.38$    &   0.93 $\pm 0.47$
 &            1.4 $\pm 0.7$ \\
 $\rm v_{CO}$ [\kms] :   & 6010$\pm 32$ & 6719$\pm 41$ & 6930$\pm 33$ \\
 $\rm W20_{CO}$ [\kms]$^{(c)}$ :   & 76    & 132 & 112  \\ 
$\rm f_{HI}$ [Jy~km~s$^{-1}]^{(d)}$ :  & 0.46 $\pm 0.07$ & 0.65 $\pm 0.10$ & ---   \\
 $\rm M_{HI}\; [10^9\times M_\sun]$ :     &       0.95$\pm 0.14$   &  1.35$\pm 0.20$ & --- \\
 $\rm v_{HI}$ [\kms] :   & 6020$\pm 31$    & 6645$\pm 63$ & --- \\
 $\rm W20_{HI}$ [\kms]$^{(e)}$ :   & 129    & 280 & ---  \\ 
SFR $\rm [M_\sun\; yr^{-1} ]^{(f)}$ : &             0.37 &      0.98 &            0.07 \\
$\rm t_{dep}\; [10^8\; yr]^{(g)}$ : &                 2.02 &              0.95 &             19.5 \\
\noalign{\smallskip} 
\hline \noalign{\medskip}
\end{tabular}
}
}

{\small
\noindent{\bf Notes:} 
$^{(a)}$  Integrated CO(2-1) flux. A 10\% calibration uncertainty is assumed which dominates the errors.
$^{(b)}$  Molecular gas mass  estimated from the  CO (2-1) data 
assuming the CO conversion factor
$\alpha_{CO10} = 1.0 $ $\rm M_\sun (K\; km\; s^{-1} pc^2)^{-1}$
and the CO(2-1) to CO(1-0) luminosity ratio
$\rm r_{21} = L'_{CO(2-1)} /  L'_{CO (1-0)} = 0.34$ \citep{Emonts2025}. A 50\% uncertainty
is assumed for $\alpha_{CO10}$ which dominates the errors.
$^{(c)}$ CO(2-1) line width measured at 20\% of peak.
$^{(d)}$  Integrated HI flux. A 15\% calibration uncertainty is assumed which dominates the errors.
$^{(e)}$ HI line width measured at 20\% of peak.
$^{(f)}$ The star formation rate (SFR)
estimated using the extinction-corrected  $\rm H_\alpha$
luminosities of the $\rm H_{II}$ regions (including only SF dominated regions and excluding the composite regions) in the SQ-A region \citep{DuartePuertas2021}.
$^{(g)}$ Depletion time scale of molecular gas: $\rm t_{dep} = M_{mol} / SFR $.
}
 \label{tab:cold_gas}
\end{table}

\begin{figure*}[!htb]
\centering
\includegraphics[width=0.8\textwidth]{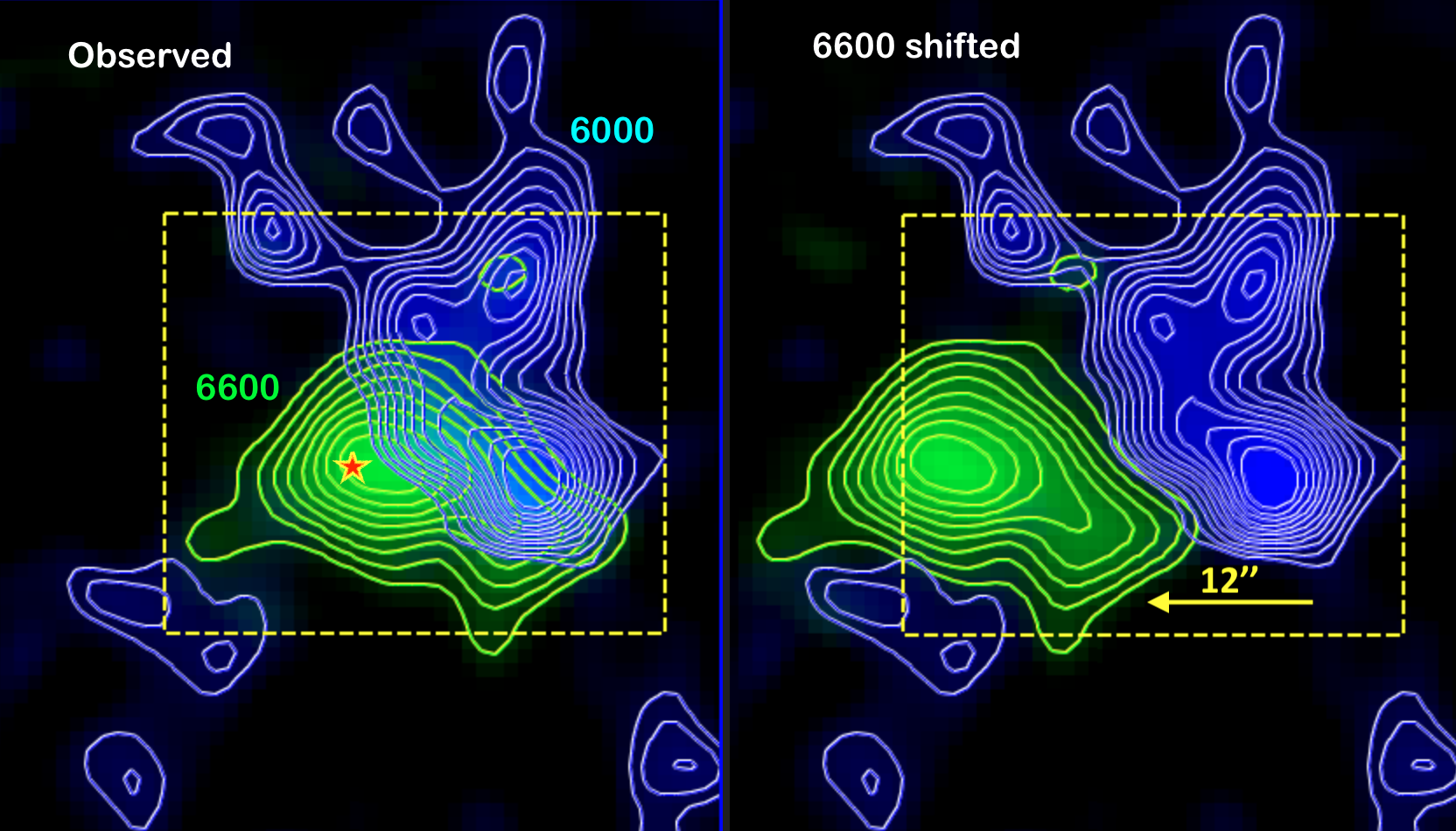}
\caption{
  {\it Left}: HI maps in the SQ-A region. The blue and green colors are for the 6000 and 6600 components, respectively. The contour levels are: 0.04, 0.05, 0.06, … Jy~beam$^{-1}$~\kms.
The red star shows the location of the SQ-A starburst (redshift: 6680 \kms, \citealt{DuartePuertas2019}).
  {\it Right}: Same as the left panel except for the map of the 6600 component is shifted eastward by 12 arcsec.
}
\label{fig:HI_shift}
\end{figure*}

The left panel of Fig.~\ref{fig:ACA_HI} shows the spatial distributions of the three
kinematic components of {\bf CO} using contours of their ACA maps, overlaid on a
JWST 15 $\mu m$ image.  The 6900 and 6000 components do not overlap
with each other but are bridged by the the 6600 component.
Many SF regions, including the SQ-A starburst that corresponds to the brightest
15 $\mu m$ peak near the  figure center, are associated with the {\bf CO gas}.
However, several SF regions in the upper-right
  corner of the figure (near the edge of the ACA coverage as shown by the
  purple dashed line), which are associated with the 6000 component
  \citep{DuartePuertas2019},
  are not significantly detected by the ACA observations because their CO emission
  is only on the 2-$\sigma$ level  ($\sigma = 0.39$~Jy~beam$^{-1}$~\kms).

The right panel of Fig.~\ref{fig:ACA_HI} presents the HI gas distributions of only the 
6000 and 6600 components while the 6900 component is missed by the HI observations.
The HI distributions of both
components are more extended than their CO counterparts. Very interestingly,
the 6600 HI seems to be located in a gap of the much more extended 6000 
HI gas. Fig.~\ref{fig:HI_shift} demonstrates that, when the 6600 HI map is shifted
relative to the 6000 HI map eastwardly by $12''$ (corresponding to a linear scale of
$\sim 5$~kpc for a distance of 94 Mpc),
the maps of the two components become very ``complementary'' in the sense that
the 6600 map fits nearly perfectly to the gap in the 6000 map.
Note that the $\sim 5$~kpc transverse displacement, without which the complementary
nature of the distributions of the two components is not easily recognizable, is the projection of a 3-dimensional relative movement
  between the two components. This is consistent with the
CCC scenario (see e.g. Fig.~8 of \citealt{Tsuge2021b}). This new piece of supporting evidence enables us to conclude with high confidence that the multiple kinematic components in SQ-A are colliding with each other and their collocation in the region is not merely a projection effect.

\citet{Fukui2021} have shown that there are two key common properties
in all cases of CCC triggered SF: (1) gas systems involved in the
collision show complementary distributions, with relative displacement explainable by relative motion and projection effects (as in the case shown in Fig.~\ref{fig:HI_shift}), and (2) a ``bridge'' (an intermediate velocity system) lies in between the
colliding systems. It appears that kinematic components in SQ-A have both of these properties (Fig.~\ref{fig:ACA_HI} and Fig.~\ref{fig:HI_shift}).
This suggests the following scenario: The SQ-A starburst is
triggered by a collision between the 6900 component 
(associated with the IGrM) and the 6000 component (associated with the intruder galaxy NGC~7318b). It appears that only a part of the 6900 component as well as
a part of the 6000 component are involved in the collision. During the collision the
6900 gas, which is likely to be denser
and more compact than the 6000 gas, is decelerated and compressed, and this decelerated and
compressed gas becomes the 6600 component in which the SQ-A starburst ($\rm v \sim 6680$ \kms)
is triggered. At the same time, the 6000 gas involved in the collision is swept into 
the 6600 component and/or dispersed, leaving a
gap in the 6000 map at  the place where the 6600 component crashes through.
The displacement shown in Fig.~\ref{fig:HI_shift}, which is $\sim 5$~kpc
in linear scale, suggests an age of the collision of a few $10^7$ years (assuming
a transverse relative velocity of a few 100~\kms between the colliding systems),
consistent with the estimated time scale for the age of the encounter between the intruder
and SQ \citep{Sulentic2001}. We will discuss this scenario in more depth 
in Section~\ref{sec:ccc}.

\begin{figure*}[!htb]
  \centering
\includegraphics[width=0.8\textwidth]{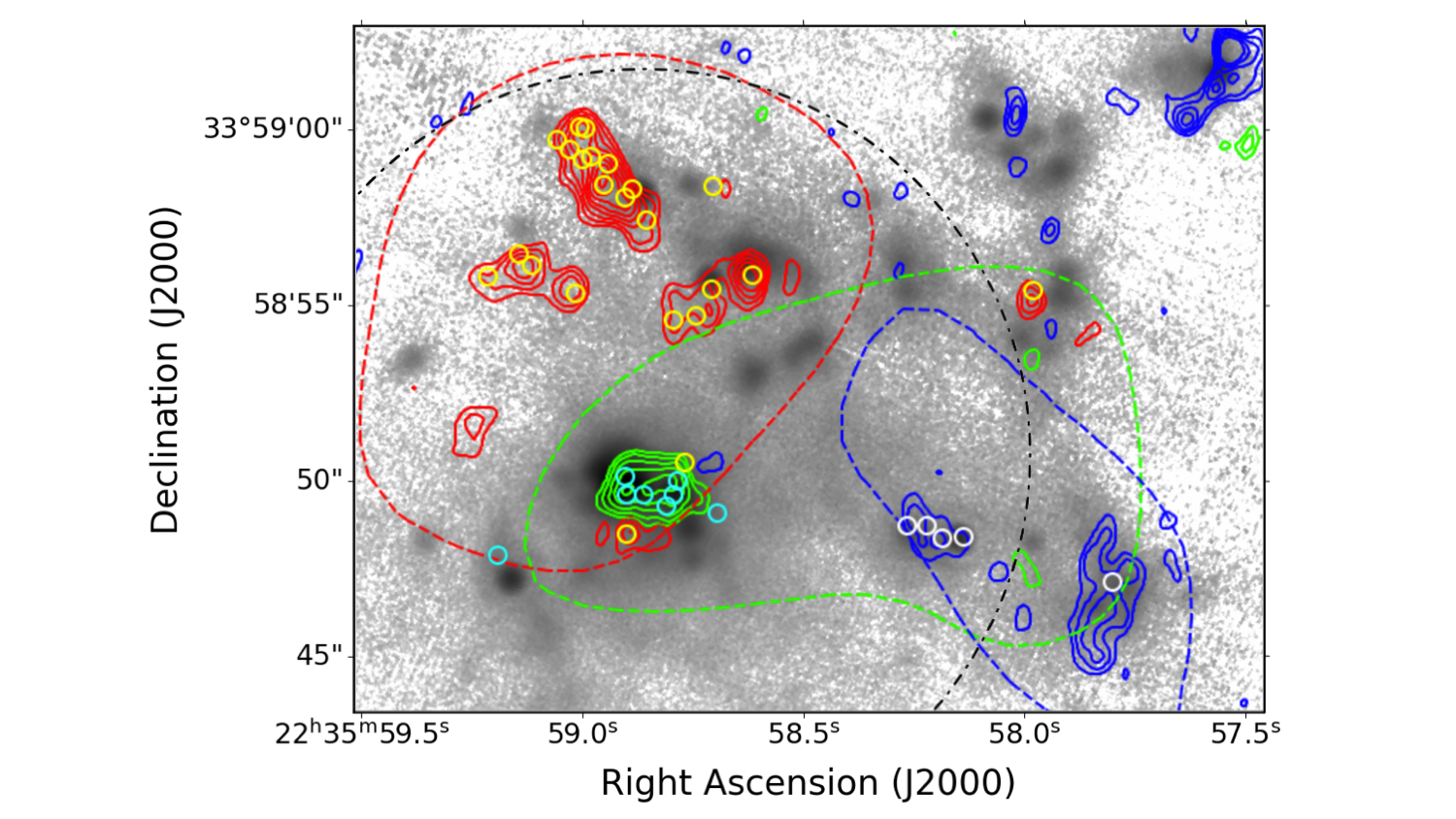}
\caption{
  Contour maps of the CO (2-1) emissions overlaid on a JWST 15 $\mu m$ image. Solid contours are for the combined TM2+ACA data which have a synthetic beam of $1.0'' \times 0.66''$, with the blue contours representing the 6000 component (levels: 0.14, 0.19, 0.24, 0.29 and 0.34 Jy~beam$^{-1}$~\kms), the green contours the 6600 component  (levels of 0.16, 0.2, 0.25, 0.3 and 0.35 Jy~beam$^{-1}$~\kms), and the red contours the 6900 component (levels of 0.22, 0.27, 0.32 and 0.37, … Jy~beam$^{-1}$~\kms). These maps have the r.m.s. of 0.048, 0.053 and 0.059 Jy~beam$^{-1}$~\kms, respectively.  Dashed contours (just one level at 1.3
  Jy~beam$^{-1}$~\kms) are for the ACA data which have a synthetic beam of $8.0'' \times 7.0''$, again the blue, green and red contours are for the 6000, 6600 and 6900 components, respectively.  The small circles ($\rm r=0.25''$) represent the clumps identified in the Field 4 of the high resolution CO (2-1) observations ($\rm beam=0.38''\times 0.23''$), with the colors of white, cyan and yellow representing the 6000, 6600 and 6900 components, respectively. The black dash-dot line marks the boundary of the primary beam ($\rm r = 11''$) of the high resolution  CO (2-1) single-pointing observation.
}
\label{fig:TM2}
\end{figure*}

In Table~\ref{tab:cold_gas}, we present the properties of the three
kinematic components.  It appears that, among them, the 6600 component
has the broadest CO and HI line widths, consistent with the hypothesis
that it is formed in the post-shock gas after the collision between
the other two components. The SFR data are adopted from
\citet{DuartePuertas2021} after adjusting the difference between the adopted
distances of SQ (they assume a distance of 88.6 Mpc).
 The total SFR of SQ and the SFR of SQ-A estimated from extinction-corrected
$\rm H_\alpha$ luminosities of the $\rm H_{II}$ regions are both consistent with the
UV-and-IR estimated values of \citet{Xu2005}. In addition, \citet{DuartePuertas2021}
provided the velocity information for the SF regions and
distinguished composite regions (shock excited) from SF regions using the BPT
diagrams. The 6600 component has the highest SFR, dominated by the starburst. 
The 6000 component has also a sizable SFR though
about a factor of 3 lower than that of 6600 component.
The 6900
component has very little SF, its SFR is more than an order of
magnitude lower than that of 6600 component. 

\section{Properties of Clumps of Molecular Gas in SQ-A}\label{sec:clumps}

\subsection{Clumps Extracted from High Resolution CO(2-1) Datacube}
In order to probe at the cloud level the physical conditions that lead to the conspicuous difference in SF activity in different kinematic components (Table~\ref{tab:cold_gas}), in this section we study ``clumps'' extracted
from the high resolution CO(2-1) datacube in the Field 4 \citep{Appleton2023}, exploiting
the algorithm ``Fellwalker'' developed by \citet{Berry2015}.
According to \citet{Li2020}, Fellwalker
shows high completeness and accuracy compared to other clump
identification algorithms. 
Fellwalker works as follows: Firstly, it selects in a 3-D
cube all pixels that are above a certain threshold, which we set at
1-$\sigma$ above the background level. Secondly, it links each of
these pixels to a local maximum along the steepest gradient.  Finally
it identifies each individual clump as the collection of all pixels that
end up at a same maximum.
Details of the clump identification procedure and definitions of
various parameters of the clumps are presented in Appendix~\ref{sec:A1}.

\begin{figure*}[!htb]
\centering
\includegraphics[width=0.45\textwidth]{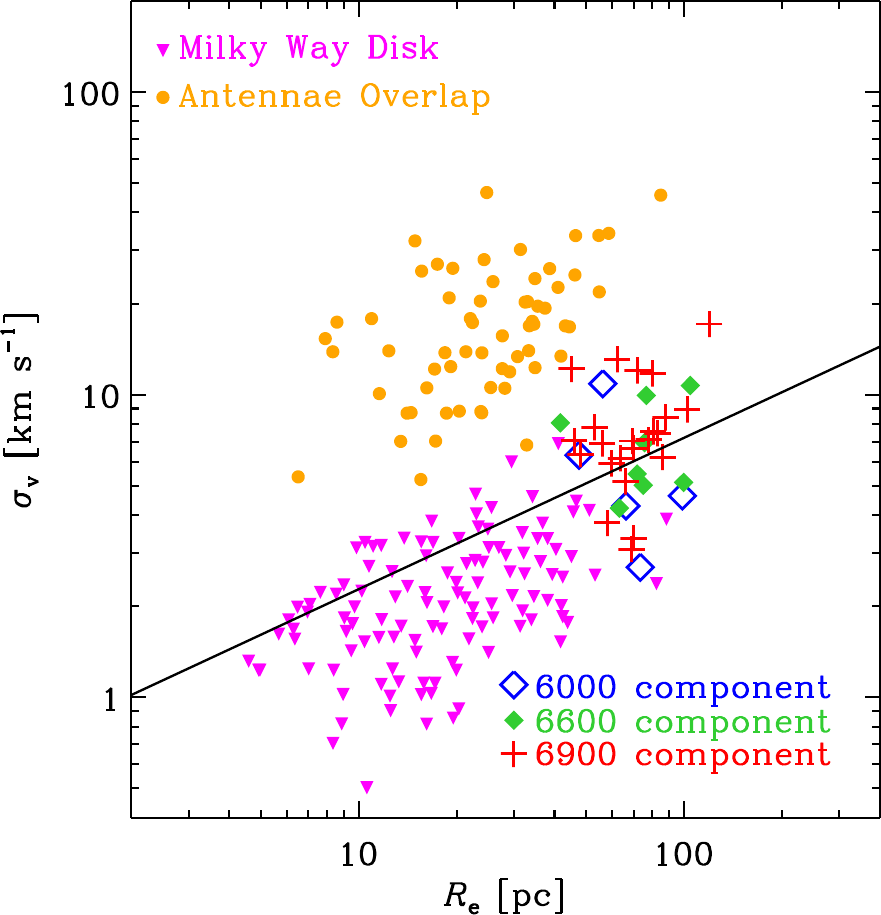}
\includegraphics[width=0.45\textwidth]{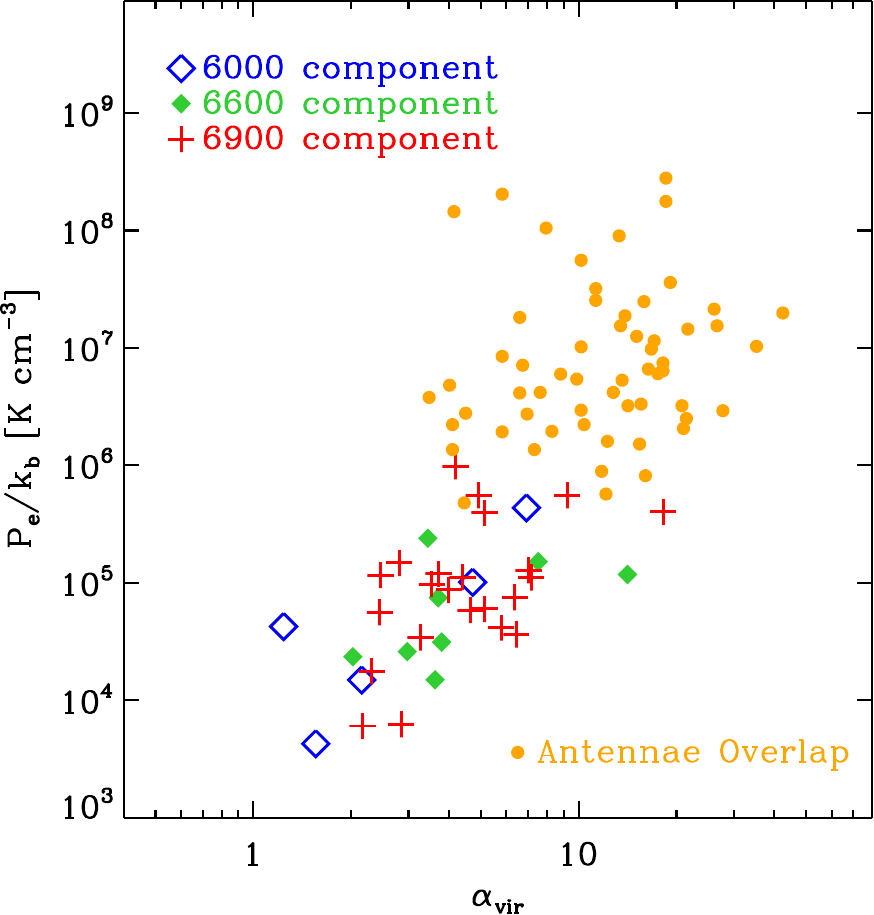}
\caption{
  {\it Left}:
  Plot of $\rm \sigma_v$ vs. $R_e$ for clumps in SQ-A region, compared with the clumps in the Antennae overlap \citep{Krahm2024} and clouds in Milky Way disk \citep{Heyer2009}. The solid line shows the typical relation for Galactic GMCs \citep{Solomon1987}.
  {\it Right}: $\rm P_e$ vs. $\rm \alpha_{vir}$ plot of clumps in SQ-A region, 
  where $\rm P_e$ is  the external pressure and $\rm \alpha_{vir}$ the virial parameter.
  The yellow dots represent clumps in the Antennae overlap \citep{Krahm2024}. 
}
\label{fig:Larson_Pe}
\end{figure*}

In total 36 clumps are identified in Field 4.
They are listed in Table~\ref{tab:clumps} and plotted
in Fig.~\ref{fig:TM2}. The figure also shows contour maps of CO (2-1) overlaid on a JWST 15 $\mu m$ image. Solid contours are for the combined TM2+ACA data ($\rm beam =1.0'' \times 0.66''$) and dashed contours mark the boundaries of ACA detections ($\rm beam =8.0'' \times 7.0''$). There are good correspondences between 
locations of the clumps and peaks of the intermediate resolution CO map (TM2+ACA) for the 6600 and 6900 components. For  the 6000 component in the north-western part 
no clumps are found because this region is outside the primary beam of the
Field 4 (whose  boundary is marked by the black dash-dot line)
and the sensitivity becomes too low. Around the SQ-A starburst (v=6680 \kms),
there is a cluster of 6600 clumps squeezed inside a compact configuration in
the TM2+ACA map. Interestingly, this 6600 configuration is surrounded by a couple
of low level contours of the 6900 and 6000 components as well as two 6900 clumps,
consistent with the hypothesis that there is an on-going interaction among
these components at this location. A couple of 15 $\mu m$ emission regions
near the figure center are not associated with clumps or peaks in the intermediate resolution CO map.
They correspond to the composite regions 
\#126 (R.A.=$\rm 22^h35^m58.6^s$,
Dec.=$+33^d58^m56.0^s$, v=6881~\kms)
and \#131 (R.A.=$\rm 22^h35^m58.6^s$,
Dec.=$+33^d58^m54.4^s$, v=6856~\kms) in \citet{DuartePuertas2019}, 
both having broad H$_\alpha$ lines and line ratios consistent with shock excitation
\citep{DuartePuertas2019, DuartePuertas2021}. The 15 $\mu m$ emission
asscociated with them must be unrelated to SF.
\begin{figure}[!htb]
\centering
\includegraphics[width=0.4\textwidth]{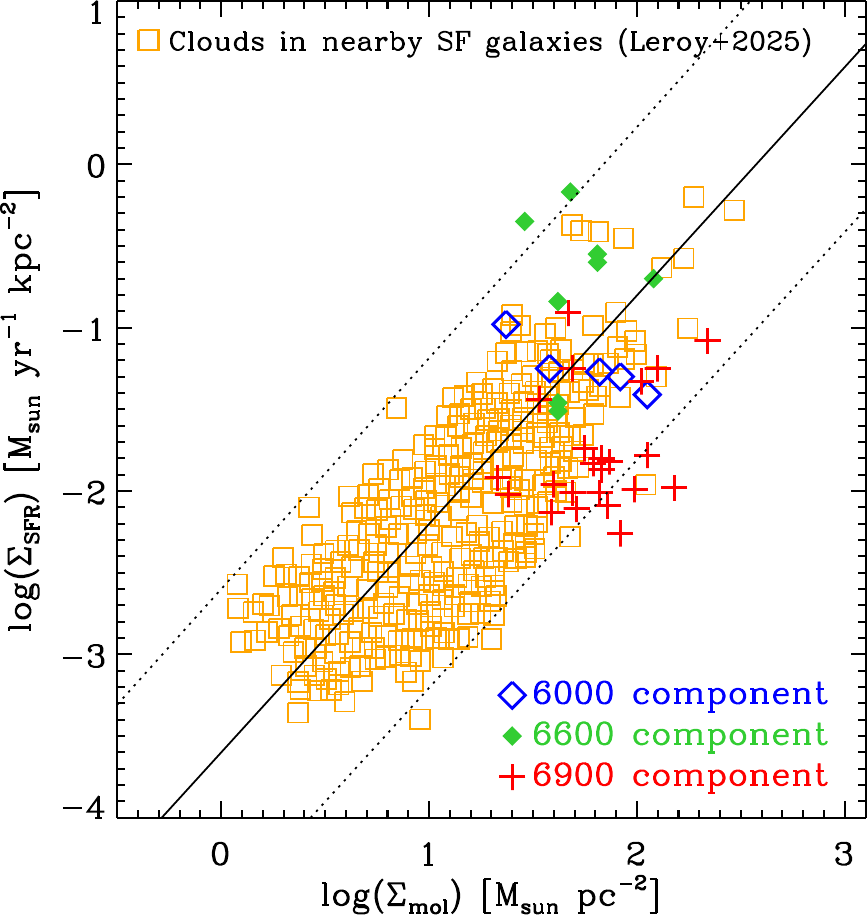}
\caption{
  The Kenicutt-Schmidt plot of the clumps in SQ-A region, compared with molecular clouds in some nearby SF galaxies \citep{2025arXiv250204481L}.
The solid line (with a slope of 1.4) represents the trend followed by
galaxies \citep{Alatalo2014} and the two dotted lines the
$\pm 1$ order of magnitude deviations. 
}
\label{fig:KS_plot}
\end{figure}

 \begin{figure*}[!htb]
\centering
\includegraphics[width=1.0\textwidth]{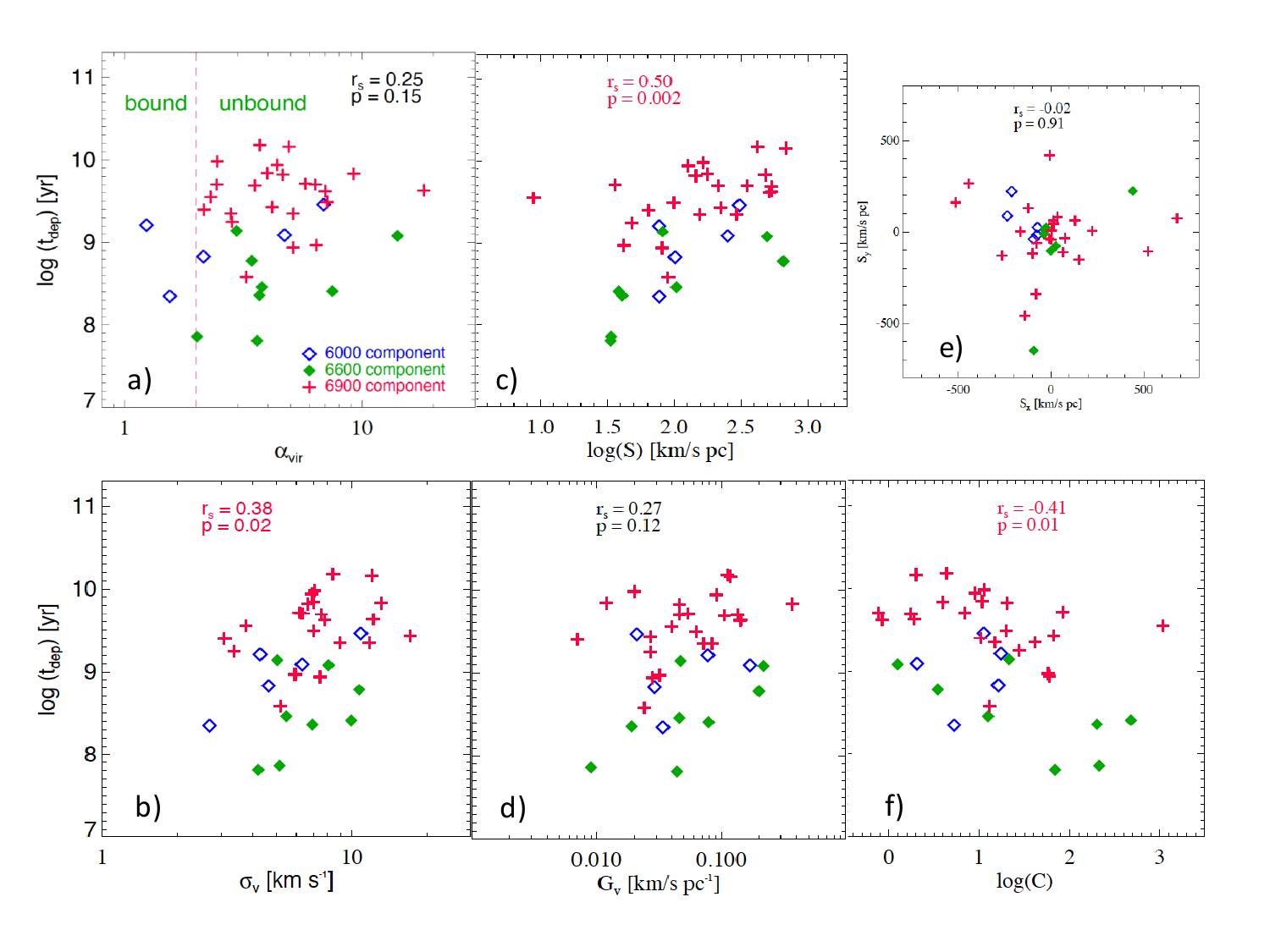}
\caption{
{\it Panel a}:  Plot of depletion time ($\rm  t_{dep}$) vs virial parameter ($\rm \alpha_{vir}$).The red dashed vertical  line at $\rm \alpha_{vir}=2$ marks the boundary separating gravitational bound and unbound clumps.
{\it Panel b}: Plot of depletion time ($\rm  t_{dep}$) vs the velocity dispersion ($\rm  \sigma_{v}$).
{\it Panel c}: Plot of $\rm  t_{dep}$ vs S parameter (the absolute value of the projected  specific angular momentum vector $\rm \vec{S}$) .
{\it Panel d}: Plot of depletion time ($\rm  t_{dep}$) vs the radial velocity gradient ($\rm  G_{v}$).  
{\it Panel e}: Plot of $\rm Sy$ (the y component  of the projected  specific angular momentum vector $\rm \vec{S}$) vs the $\rm S_x$ (the x component of the vector). It shows
no trend for a preferred orientation of the vector.
{\it Panel f}: Plot of $\rm  t_{dep}$ vs C parameter (the nonrotational-to-rotational energy ratio).
The Spearman rank correlation coefficient $\rm r_s$ and the significance parameter p (the probability of the null hypothesis) are listed in each plot, with red characters indicating significant ($\rm p \leq 0.01$) or
marginally significant correlations ($\rm 0.01 < p \leq 0.03$). 
}
\label{fig:fig8}
\end{figure*}

\subsection{Dynamical Properties of Clumps}
 The left panel of Fig.~\ref{fig:Larson_Pe} shows the plot of velocity dispersion
 $\sigma_v$ (Eq.~\ref{Eq:sigma_v})
 vs. the effective radius $\rm R_e$  (Eq.~\ref{Eq:Re})
 of the clumps.
They have radii in the range of
   45 -- 100 pc (with a mean of 72$\pm 3$~pc)
   and velocity dispersions in the range of
  2 -- 20~\kms (mean = $7.4 \pm 0.5$~\kms).
It is worth noting that quite a few clumps have sizes ($\rm = 2\times R_e$) smaller than the spatial resolution of the ALMA observation ($\sim 120$~pc). This is because
$\rm  R_e$ is derived from an effective area $\rm A_e$ which in turn is a flux-weighted summary of all pixels in a clump  (Eq.~\ref{Eq:Ae}). The radius so defined has the advantage of being insensitive to the clump selection effects, but is vulnerable to the error of the flux (similar to the error in measurement of the beam size using a point source image). For faint clumps with $\rm f_{CO21}\leq 0.03$~Jy~\kms, corresponding to log($\rm M_{mol}/$\msun )$\leq 5.7$, the error of $\rm R_e$ is on the order of 25\%.

 There is no significant difference in either
$\rm R_e$ or $\sigma_v$ between the three components. When compared
with the clouds in the Milky Way disk \citep{Heyer2009}, the SQ-A
clumps are in the regime of very large giant molecular clouds
(GMCs). Indeed they fall right on the line of the so called "Larson
Law" (the solid line in the plot) found for the Galactic GMCs
\citep{Solomon1987}. On the other hand, the clouds in the overlap region of the Antennae
Galaxies (hereafter the Antennae overlap)
have significantly higher velocity dispersions, $\sim 5\times$
above those of SQ-A clumps and clouds in the Milky Way disk,
indicating different physical conditions in them \citep{Krahm2024}.

The right panel of Fig.~\ref{fig:Larson_Pe} is a plot of the
external pressure $\rm P_e$ (Eq.~\ref{Eq:Pe}) vs. the virial parameter
$\rm \alpha_{vir}$ (Eq.~\ref{Eq:avir}).  Here again the clumps in different
kinematic components are well mixed with each other.  All but two of 
them have $\rm \alpha_{vir} > 2$, namely are unbound by gravity.
The $\rm P_e/k_B$ is in the range of $3\times 10^3$ -- 10$^6$ K~cm$^{-3}$,
significantly lower than that of clumps in the Antennae overlap
\citep{Krahm2024}. Rather, the $\rm P_e/k_B$ of SQ-A clumps
is similar to that of the GMCs in the Milk Way and nearby galaxies
\citep{Heyer2009, Donovan2013}.  Note that adopting a higher CO
conversion factor would increase the value of $\rm P_e/k_B$ and reduce
the value $\rm \alpha_{vir}$. There is a strong correlation between
$\rm P_e/k_B$ and $\rm \alpha_{vir}$ (the Spearman correlation
coefficient $\rm r_s=0.62$ with a significance of p$=6 \times
10^{-5}$), indicating that the clumps are mainly confined by the
external pressure instead of the self-gravity. The clouds in the Antennae
overlap, which have much higher
$\rm P_e/k_B$ and  $\rm \alpha_{vir}$ values,
lie slightly (by $\sim 0.5$ dex) above this correlation.

\subsection{Star Formation Properties of Clumps}
Fig.~\ref{fig:KS_plot} presents the $\rm \Sigma_{SFR}$ vs. $\rm \Sigma_{mol}$
plot (the ``Kennicutt-Schmidt plot") of the clumps. The solid line (with a slope of 1.4)  in the figure represents the trend followed by galaxies \citep{Alatalo2014} and the two dotted lines the
$\pm 1$ order of magnitude deviations. Clumps of three different
kinematic components have similar $\rm \Sigma_{mol}$ values.  In contrast,
their $\rm \Sigma_{SFR}$ values are stratified over  $\sim 2$ orders of magnitude range: 6600 clumps are on the top, 6000 clumps in the
middle, and 6900 clumps at the bottom. The high SFR enhancement of 6600
clumps is consistent with the nature of the SQ-A starburst which is associated
with this kinematic component. The 6000
clumps are also undergoing active SF, while most of 6900 clumps are
clearly suppressed with the $\rm \Sigma_{SFR}$ about an order of magnitude
lower than the nominal line. In SQ-A clumps, $\rm \Sigma_{SFR}$ does not depend on
$\rm \Sigma_{mol}$ at all: their Spearman rank correlation
coefficient is $\rm r_s = -0.04$ and the significance parameter
(i.e. probability of the null hypothesis) $\rm p=0.82$.

The depletion time scale $\rm t_{dep}$, the
inverse of the SFR per unit $\rm M_{mol}$, is another SF activity indicator.
Compared to $\rm
\Sigma_{SFR}$, $\rm t_{dep}$ has the advantage of being
an intensive physical property and therefore
insensitive to the 'rich man effect' (large clouds having large amount of
everything). Panel-a of Fig.~\ref{fig:fig8} presents a plot of
$\rm t_{dep}$ vs the virial parameter $\rm \alpha_{vir}$. There is no
significant correlation between these two variables, another
evidence for the self-gravity not playing an important role in
regulating the SF activity of the clumps. On the other hand, there
is a marginally significant correlation 
between $\rm t_{dep}$ and the
velocity dispersion $\rm \sigma_v$
(the significance parameter p=0.02, panel-b of Fig.~\ref{fig:fig8}).
Given the fact that the clumps have $\rm \sigma_v$ much larger than
the sound speed ($\rm v_{sound} \sim 0.2$~\kms for gas of $\rm T \sim
10$~K) and therefore are dominated by non-thermal motions,
the correlation suggests that the gas motion may play a significant
role affecting the SF activity in the clumps.

The non-thermal motions can be in many different forms, including
contraction, expansion, rotation, shear, and small scale turbulent
motions. Together with the thermal motion, they form a
hierarchy and the energy cascades down from large scales to small
scales.  In collision systems, the main energy source is the shock
triggered by collision \citep{Guillard2009, Guillard2022}. With a
resolution of $\sim 120$~pc, we cannot resolve the small scale
turbulence. Therefore we begin by checking the effects of rotation and
shear on $\rm t_{dep}$. A plot of $\rm t_{dep}$ vs the S parameter (Eq.~\ref{Eq:S})
is presented in the panel-c of Fig.~\ref{fig:fig8}. Being a measure
of the specific angular momentum, the S parameter is
related to a centrifugal force that can balance gravity and suppress
the SF.  Indeed we see a significant correlation between $\rm t_{dep}$
and S with a Spearman correlation coefficient of $\rm r_s = 0.50$ and
a significance of p=$2 \times 10^{-3}$.
There is a trend that kinematic component with higher average
$\rm t_{dep}$ (i.e. low SF activity) has also higher average S value
($\rm \langle log(S) \rangle = 1.96\pm 0.19, \; 2.13\pm 0.13, \; 2.19\pm 0.10$ for
the 6600, 6000 and 6900 components, respectively). The S values 
have large scatters within individual components and clear dependence
  of $\rm t_{dep}$ on S can be recognized in each of all three components.
 It should also be noted that the $\rm t_{dep}$ vs S
plot is still stratified in the sense that, for a given S, the 6900
clumps tend to have high $\rm t_{dep}$ and the 6600 clumps low $\rm
t_{dep}$ and the 6000 clumps in between. This indicates that some other factors, which
may be systematically different in the different kinematic components, can significantly
affect the SF in clumps.
\citet{Braine2018,
  Braine2020} found a tendency for rotation vectors of molecular
clouds in M33 and M51 to align with the galactic disk
rotation. However,
for SQ-A clumps no trend is found for the orientation of the 2-D
vector $\rm \vec{S}$ (panel-e of Fig.~\ref{fig:fig8}).
This is consistent with the rotation being
driven by localized torques (presumably associated with the collision)
instead of effects associated with large scale motions
(as in the cases of M33 and M51 clouds).

Local torque can be generated by shear motion inside the clump, which in
turn can be powered by a shock (particularly an oblique shock).
The amplitude of a shear can be measured by velocity gradient.
Panel-d of Fig.~\ref{fig:fig8} presents a plot of
$\rm t_{dep}$ vs the radial velocity gradient $\rm G_{v}$ (Eq.~\ref{Eq:Gv}), which shows
no significant correlation between the two variables.
This suggests that shear motion may not have a
significant direct impact to the SFR.

Panel-f of Fig.~\ref{fig:fig8} presents a plot of $\rm t_{dep}$ vs
the C parameter, the nonrotational-to-rotational energy ratio (Eq.~\ref{Eq:C}).
There is a significant anti-correlation in this plot,
with the Spearman correlation coefficient of $\rm r_s = -0.41$ and
significance $\rm p= 0.01$.  Here we see an even better separation in
C than in the S parameter between the three kinematic components: on average
the 6600 component has the highest C and lowest $\rm t_{dep}$ and the
6900 component the lowest C and highest $\rm t_{dep}$.
While the correlation between $\rm t_{dep}$ and the S
parameter indicates a role of rotation
in suppressing the SFR, the anti-correlation between
$\rm t_{dep}$ and the C parameter suggests a positive role for non-rotational motion
in enhancing the SFR. Simulations have shown that collisions can generate
compressive turbulence \citep{Takahira2014} which in turn
can enhance the SFR \citep{Krumholz2014, Renaud2014}. It is plausible that
the compressive turbulence, which is included in the non-rotational motion, is
the engine that drives the enhanced SFR.

By measuring the rotational support and non-rotational motions at cloud scale, we can estimate the degree of stability of the molecular structures, and connect it to the star formation activity. However, this reasoning is not sufficient to explain the starburst nature of these regions, that is why star formation proceeds faster here than in classical clouds of the same molecular gas mass. \citet{Renaud2014} used simulations of interacting galaxies to establish that the starburst regime is reached when the interstellar turbulence becomes dominantly compressive (as opposed to solenoidal) as the result of galactic-scale tidal compression. This induces an excess of dense gas which not only allows for elevated star formation (high SFR) but also fast star formation (short depletion time). The same effect could well be active in SQ-A, with compression originating from the interaction between the 
intruder-associated material and the IGrM, and its energy being converted to compressive turbulence, in turn triggering the observed starburst.
{While our result on the anti-correlation between $\rm t_{dep}$ and the C parameter  seems to support this hypothesis, it is important to note that we do not have any direct
measure of the turbulent motion. Indeed, the ultimate test of this hypothesis would require the measurement of turbulent motions within the clouds themselves, which remains out of reach of the present capacities. 

It should also be noted that all the correlations presented in Fig.~\ref{fig:fig8}
are quite scattered even when they are significant ($\rm p \leq 0.01$)
or marginally significant ($\rm 0.01 < p \leq 0.03$). 
Although there is a trend that clumps in the more SF active component
tend to have lower S parameters
and higher C parameters (e.g. $\rm \langle log(S) \rangle = 1.96\pm 0.19$
and  $\rm \langle log(C) \rangle = 1.53\pm 0.32$ for those in 6600 component),
the differences with those in low SF activity components
(e.g. in the 6900 component $\rm \langle log(S) \rangle = 2.19\pm 0.10$ and
$\rm \langle log(C) \rangle = 1.09\pm 0.15$) are
only at 1-$\sigma$ level between the corresponding means.
The large scatters within the same kinematic component are
understandable because many factors
which are not controlled in this analysis
can affect the variables. These include the variation of the evolution
stages among the clumps, which strongly affects the estimate of
$\rm t_{dep}$ \citep{Kruijssen2018, Kim2022}. The projection effect and beam smearing effect
can affect many parameters such as
$\rm R_e$, $\rm \sigma_v$, S parameter (projected specific angular
momentum), C parameter (nonrotation-to-rotation energy ratio), and
the velocity gradient ($\rm G_v$ parameter). The small scale variation of the CO conversion factor,
particularly if it is $\rm t_{dep}$ dependent \citep{Renaud2019a},
can  also add on to the scattering of $\rm t_{dep}$ significantly.
Nevertheless, the correlations
and non-correlations between $\rm t_{dep}$ and various parameters may provide
important insights to the factors that control the SF activity at the cloud level. In
particular the significant correlation between S and $\rm t_{dep}$ strongly suggests
that the rotation may play an important role in hindering collapse
and suppressing the SFR.
At the same time, our finding that the  $\rm t_{dep}$ distribution is still stratified among different components for a given S (Fig.~8c) demands other factors which can affect the triggered SF in clumps, such as the compression of gas hinted at by the significant anti-correlation between C and $\rm t_{dep}$.

\section{Discussion}\label{sec:discussion}

\subsection{CCC Model and SQ-A Starburst}\label{sec:ccc} 
Fig.~\ref{fig:ACA_HI} and Fig.~\ref{fig:HI_shift} provide new
supporting evidence for the hypothesis that the SQ-A starburst is
triggered by a collision between two gas systems, one associated
with the IGrM ($\rm v \sim 6900$~\kms) and another with the intruder galaxy
NGC~7318b ($\rm v \sim 6000$~\kms) .
Here we discuss this hypothesis in the context of the
CCC theory. Originally, the CCC theory was proposed to interpret the
triggered SF in Galactic GMCs, particularly the formation of young massive
 star clusters (YMCs; \citealt{Habe1992, Inoue2013,
  Takahira2014}). More recently, this theory was applied to
starbursts in large extra-galactic regions (size up to a few kpc)
involved in galaxy-galaxy collisions, such as the overlap region of the Antennae Galaxies \citep{Tsuge2021a, Tsuge2021b} and 30 Doradus plus southern ridge of the
LMC \citep{Fukui2017, Tsuge2024}. Similar to SQ-A,
the Antennae overlap region is undergoing high SF enhancement due to a
direct collision of two gas systems associated with two
galaxies, respectively \citep{Whitmore1995, Xu2000}. However, there are remarkable differences
between the two cases.  As shown in Table~\ref{tab:cold_gas}, the
kinematic systems in SQ-A are dominated by the HI gas (the
$\rm M_{HI}$-to-$\rm M_{mol}$ ratio $\sim 10$). And the complementary
distributions among the
different components are found only in the HI maps, not in the CO
maps (Fig.~\ref{fig:ACA_HI}). On the other hand, the collision in the
Antennae overlap is
happening primarily in the molecular gas which is much more massive
than the HI \citep{Wilson2000, Hibbard2001}, and the complementary
distributions of the colliding systems are found in the CO maps
\citep{Tsuge2021a}. Also, clouds in the
Antennae overlap have much higher velocity dispersions and external pressures, 
$\sim 5\times$ and $\sim$two orders of magnitude higher than those in SQ-A clumps, respectively (right panel of Fig.~\ref{fig:Larson_Pe}). Since the collision velocity in SQ-A is much higher than that in the Antennae overlap which is $\sim 100$~\kms \citep{Tsuge2021a}, the low pressure in the former is suggestive of a much lower pre-collision density than in the latter, given the relation $\rm P_e \sim \rho_{pre} v_{coll}^2$ where $\rm \rho_{pre}$ is the pre-collision density and $\rm v_{coll}$ the collision velocity. 

Indeed, SQ-A may be more comparable to the giant
SF regions in 30 Doradus which are produced presumably by a collision of
two HI systems, one associated with the LMC and the other accreted from
the Small Magellanic Cloud (SMC; \citealt{Fukui2017, Tsuge2024}).  
It is worth noting that,
similar to the LMC, SQ-A has a sub-solar metallicty of $\rm Z \sim 0.5\;
Z_\sun$ \citep{DuartePuertas2021}, which might be the reason for its
high $\rm M_{HI}$-to-$\rm M_{mol}$  ratio. In direct GMC-GMC collisions, SF is
triggered in the denser and smaller cloud when it crashes into a
less dense and larger cloud \citep{Habe1992, Fukui2018}.  On the
other hand, the triggered SF in HI-HI collision is different.  As
shown in the simulations of \citet{Maeda2021, Maeda2024}, high
velocity collisions ($\rm v_c \gsim 50$~\kms) of HI gas can trigger the
formation of a large amount of dense cores (density
$\rm > 10^4\; cm^{-3}$) which in turn form YMCs in a short time scale
comparable to the free-fall time of the dense cores ($\rm t_{ff} \lsim 10^6$ years).  In
these models, because of the rapid cooling ($\rm t_{cool} < 10^6$
years), the density of the post-shock gas is approximately  proportional to the shock velocity which is about a half of the
collision velocity.  Assuming a collision
velocity of $\sim 1000$\kms (as it is in SQ-A) and an initial density
of $\rm n_0 = 1 \; cm^{-3}$, the post-shock gas density becomes as
high as $\rm n \sim 350\; cm^{-3}$ \citep{Maeda2024}. This gas is thermally unstable and
has a free-fall time scale of
$\rm t_{ff} \sim 2 \times10^6\; yr  \times (n/350cm^{-3})^{-0.5}$.
After a few $\rm t_{ff}$ dense cores will
form and the starburst follows. The whole process should occur within
a time scale of $10^7$ years, consistent with the time scale we found
for the collision in SQ-A. 
Results of the clump analysis in Section~\ref{sec:clumps}, which shows
a lack of dependence of SFR on density
(Fig.~\ref{fig:KS_plot}) and an anti-correlation between
$\rm t_{dep}$ and the C parameter (panel-f of Fig.~\ref{fig:fig8}), also support
the scenario in which
much of the  SF in SQ-A is collisionally triggered
instead of gravitationally induced.
\citet{Gallagher2001} and \citet{Fedotov2011} found large amount
of young stellar clusters (ages of $\leq 10^{7}$ years) in SQ-A,
another supporting evidence for collisionally triggered SF in the region
given the tight relation between young clusters and CCC \citep{Fukui2021}.
Finally, although in SQ-A the CCC mainly occurs between
HI dominated clouds (i.e. HI-HI collision), HI-GMC and GMC-GMC collisions
may also contribute to the SF enhancement albeit with relatively low significance
given the low $\rm M_{mol}$-to-$\rm M_{HI}$ ratio.

\subsection{Collision of Cloud with Diffuse Gas and Suppression of SFR}\label{sec:quenching}
A major challenge for the collisional scenario of the SQ-A starburst
is the severely suppressed SF activity in the large-scale shock region
\citep{Cluver2010, Guillard2012}. Why does a high-speed
collision in the large-scale shock region cause an SF suppression while in SQ-A a
starburst?

There is a stark difference between the collisions in the
two regions: in SQ-A both the IGrM and the intruder-associated
components have large amount of cold gas. On the other hand, in the shock
region, only the CO of molecular gas of the intruder-associated component (i.e. the
6000 component) is significantly detected \citep{Lisenfeld2002, Guillard2012,
  Emonts2025} whereas
no significant CO or HI detection is found  for the IGrM-associated component.
\citet{Sulentic2001} suggested that the
reason for the lack of cold IGrM gas in the shock region is because it
has been converted to hot X-ray gas and warm ionized gas after the collision. This
interpretation is unlikely to be true because, as shown in simulations
(e.g. \citealt{Maeda2021}), the cooling time in cloud-cloud collisions
is so short that the post-shock gas becomes cold and neutral again soon after
collision (on a timescale of $< 10^6$ yrs). It is more likely that
the pre-shock IGrM in the shock region is predominantly in the diffuse
warm neutral medium (WNM) or/and the warm ionized medium (WIM;
\citealt{Iglesias-Paramo2012, Rodriguez-Baras2014, Arnaudova2024}). 
The WNM gas is too diffuse to be detected by VLA but can be
picked-up by sensitive single-dish HI observations. Indeed recent FAST
observations of SQ have found more HI gas in the 6600 components, most likely
in the diffuse IGrM \citep{Xu2022, Cheng2023}. If indeed 
 the pre-shock IGrM gas in the shock region is diffuse and of low density,
the high speed collision between this gas and the diffuse gas
associated with the intruder can produce two blast wave shocks, one
propagates in the IGrM and the other in the system associated with the
intruder.  The shock ridge shown in the X-ray and the radio continuum
is presumably associated with the post-shock gas between these
two shock fronts.  From the
X-ray data \citet{Trinchieri2003} found a pre-shock gas density on the
order of $\rm \sim 10^{-2}\; cm^{-3}$.  Therefore, the process affecting
the clouds in the shock region may be better analyzed using
models for interaction of blast wave shock and interstellar clouds
\citep{McKee1975, Klein1994, Guillard2009} instead of CCC
models. One of the key findings of \citet{Klein1994} is that strong
vorticity can be generated at the cloud-intercloud boundary and in the
tail behind the cloud, both by the initial passage of the blast wave shock and by
the subsequent flow of post-shock inter-cloud gas passing-by the cloud.
We argue that it is the vortices so generated that suppress
the SF in molecular clouds in the shock region. 
The vortices can contribute significantly to the solenoidal
turbulence which acts against cloud contraction, and may eventually
destroy the cloud \citep{Klein1994}.
Both theoretical simulations and observations in literature have shown that,
while the compressive turbulence can enhance SFR, the solenoidal
turbulence can suppress SFR \citep{Federrath2012, 
  Orkisz2017, Rani2022, Petkova2023}.

Similar to clouds in the shock region, clumps of the 6900 component in
SQ-A  have severely suppressed SF activity.
It is possible that these clumps have avoided cloud-cloud collision, given the
fact that they have not been decelerated significantly. If these clumps
are also undergoing interactions with blast wave shocks due to
collisions between diffuse media in SQ-A, the above scenario
proposed for clouds in the shock region can apply and explain
why the SFR is suppressed in 6900 component.
The significant correlation between
$\rm t_{dep}$ and the S parameter, which indicates
an important role of rotation in suppressing SFR, supports this hypothesis.
The weak X-ray emission in SQ-A may not be
used as evidence against such an analogy because the X-ray may be affected
by absorption \citep{OSullivan2009}.
It is interesting to note that there is
  a faint stripe of diffuse X-ray emission extending to
  the east of SQ-A  \citep{OSullivan2009},
which can also be seen in the VLA 1.4 GHz radio continuum map \citep{Xu2003} and
faintly visible in the H$_\alpha$ map (Fig.~\ref{fig:fig1}).
This feature may be shock-excited, providing
further evidence for shocks triggered by the collision between diffuse media in the region
around SQ-A.

Unlike the 6000 component which is linked to
the intruder galaxy NGC~7318b by a long arm or a partial ring \citep{Emonts2025},
the origin of the 6900 component in SQ-A is still unknown. It might be related to the large CO emission region in the north-east of NGC~7319  nucleus which has  a system velocity of $\sim 6900$~\kms \citep{Yun1997, Gao2000, Emonts2025}, though no clear link has been detected between that region and SQ-A in any band. Future high angular resolution and high sensitivity HI observations covering the velocity range beyond 6800 \kms will help resolving this issue.

Our interpretation for the diversified $\rm t_{dep}$ in colliding systems in SQ can
have far-reaching inference to theories of interactinon-induced SF in galaxy mergers,
particularly those in early merging stages (e.g. optical-NIR selected pairs). Given the
nearly universal detections of young massive star clusters (YMCs) in merging galaxies
\citep{Whitmore1995, Adamo2020} and the tight relation between
YMCs and collision triggered SF \citep{Elmegreen1998, Fukui2021}, the CCC
mechanism must play an important role in the SF activity of galaxy mergers.
Our results for SQ-A indicate that effects of collisions are very complex, can cause either
SF enhancement or SF suppresion depending on the density contrast between the gas
systems involved in the collision. In studies of
a sample of K-band selected close major-merger pairs (KPAIR), it has been
found that the SF galaxies in spiral-spiral pairs have strong SF enhancement while those in spiral-elliptical
mixed pairs have not \citep{Xu2010, Cao2016, Lisenfeld2019}. Simulations predict enhancement of compressive turbulence in early mergers \citep{Renaud2014} which triggers formartion of large amount of dense cores and results in widely spread SF enhancement in the disks, consistent with the observations for extended SF enhancement
in disks in spiral-spiral pairs
\citep{Iono2005}. Why does this mechanism fail to work for SF galaxies in mixed pairs?  \citet{Xu2021} found that galaxies in mixed pairs are more likely involved in high-inclination orbit and high speed collisions, which may trigger density waves (as shown by ring-like structures) in the disk. It is plausible that the shocks associated with density waves may act similarly as the blast wave shock in the model of \citet{Klein1994} and suppress the SF of the gas clouds. As an example, the SFR/$\rm M_{mol}$ (the inverse of $\rm t_{dep}$ ) of the spiral galaxy NGC~2936 in the mixed pair Arp~142, which has gone through a high-speed and high-inclination collision recently, is significantly lower than the nominal value for SF regions \citep{Xu2021}.

\subsection{Uncertainties of CO Conversion Factor and $\rm r_{21}$}\label{sec:conversion}

In this paper, we adopt a CO conversion factor
$\rm \alpha_{CO10} = 1.0\; M_\sun (K\; km\; s^{-1} pc^2)^{-1}$.
This is a factor of 3.2 or 4.3
lower than the Galactic value, depending on whether the contribution
of the heavy elements is included \citep{Bolatto2013}.  Starburst
galaxies often have low CO conversion factors ($\rm \alpha_{CO10} <
1$; \citealt{Downes1998}). There is also evidence for molecular gas
systems involved in high speed collisions to have low
$\rm \alpha_{CO10}$.  For gas in the bridge between Taffy Galaxies,
\citet{Braine2003} and \citet{Zhu2007} found a $\rm \alpha_{CO10}$ 4
-- 10 times lower than the Galactic value. Given that SQ-A is also
involved in a high speed collision with a collision velocity comparable
to that between Taffy Galaxies, the low $\rm \alpha_{CO10}$ is
justified. On the other hand, the slightly sub-solar metallicity of SQ-A
($\rm Z \sim 0.5 Z_\sun$, \citealt{DuartePuertas2021}) may hint a higher
$\rm \alpha_{CO10}$ \citep{Bolatto2013}. Also, the average virial parameter
of the SQ-A clumps is $\alpha_{vir} \sim 3$ (see right panel of
Fig.~\ref{fig:Larson_Pe}), indicating that the virial mass is about
3 times higher than the mass estimated using the adopted
$\rm \alpha_{CO10}$. However in colliding systems and starbursts, the virial
equilibrium may be broken and virial mass may not be a good estimator of the
true mass.  Constraining the
$\rm M_{mol}$ using the local thermodynamic equilibrium (LTE) models,
\citet{Krahm2024} found that the conversion factor 
  in the Antennae overlap is consistent with
  $\rm X_{CO} = 0.5\times 10^{20} \; cm^{-2} (K\; km\; s^{-1})^{-1}$, corresponding to
  $\rm \alpha_{CO10} = 1.075\; M_\sun (K\; km\; s^{-1} pc^2)^{-1}$  \citep{Bolatto2013},
  and
that the clouds in the Antennae overlap are highly super-virial with
a mean $\alpha_{vir} = 13.77\pm 0.37$. Actually, these authors found
that molecular clouds are in general super-virial in starburst galaxies.
In summary, the CO conversion factor in SQ-A is highly uncertain. Future works
involving CO$^{13}$ observations and modelings, e.g.  using the LTE models or 
the large velocity gradient (LVG) models \citep{Bolatto2013},
may help to resolve this issue. 

Also, the adopted CO(2-1) to CO(1-0) luminosity ratio $\rm r_{21}=0.34$ is quite uncertain, too, because SQ-A is near the edge of the primary beam of the CARMA CO(1-0) observations based on which the $\rm r_{21}$ value is estimated. In addition,
the 6000 component (associated with the intruder)
may have a slightly different value ($\rm r_{21}=0.29$; \citealt{Emonts2025}). 

The uncertainties of $\rm \alpha_{CO10}$ and $\rm r_{21}$ affect the estimate of $\rm M_{mol}$ from the ALMA CO(2-1) data through a combined factor $\rm \alpha_{CO10}\times r_{21}$.  Nevertheless our main conclusions, which are primarily 
based on gas distributions and correlations (and lack of them) between various
parameters in clumps, are not significantly affected by these uncertainties. 

\section{Conclusions}
In this paper we aim to achieve two science goals:
(1) proving the validity of the hypothesis that the starburst SQ-A in
the IGrM of Stephan's
Quintet (SQ) is triggered by  a high-speed collision ($\sim 900$~\kms) 
  between two gas systems, one associated with the IGrM
  ($\rm v \sim 6900$~\kms) and another with the intruder galaxy
  NGC~7318b ($\rm v \sim 6000$~\kms);
(2) solving the puzzle of why gas clouds in SQ-A and in the large-scale shock region
  of SQ, which are both involved in the same high-speed collision between the IGrM
  and the intruder, can have dramatically different SFRs.
To achieve these goals, we
  exploit the following data sets: (1)  CO(2-1) datacubes with angular
  resolutions in the range of $\rm 0.2'' - 7.0''$ obtained using
  ALMA in various configurations (with adopted distance of 94 Mpc,
    1'' corresponds to 0.45 kpc);
  (2) an  HI datacube of angular resolution
  of $\rm 6.6'' \times 7.9''$ obtained using the VLA B, C \& D arrays;
  and (3) a JWST MIRI image in the 15 $\mu m$ band.
We reach the following conclusions:
  
\begin{description} 
\item[1] 
  The  CO(2-1) data show clearly three kinematic systems in SQ-A that correspond to
  the 6000, 6600 and 6900 components in the literature
  \citep{Shostak1984, Williams2002, Cheng2023},
  with systematic velocities of $6010\pm 32$, $6719\pm 41$ and $6930\pm 33$~\kms,
  respectively.  As shown in the CO maps covering the
entire SQ \citep{Emonts2025}, both the 6600 and 6900
components in SQ-A are most likely associated with the IGrM while the 6000 component
with the intruder galaxy NGC~7318b. The 6900
component has the highest molecular gas mass ($\rm M_{mol} = 1.4\pm 0.7\; 10^8 M_\sun$) which
is 1.9 and 1.5 times of those of the 6000 and 6600 components, respectively.
The 6900 and 6000 components do not overlap
with each other spatially but are bridged by the 6600 component.
\item[2]
The 6000 and 6600
components are also detected in new HI data with velocities of
  $6020\pm 31$ and $6645\pm 63$~\kms, respectively, 
slightly different from those of the corresponding components in the CO.
The 6900 component is not observed in the HI because it is 
  outside the frequency coverage of the VLA observations.
  Both the 6000 and 6600 components show rather high $\rm M_{HI}$-to-$\rm M_{mol}$ ratios ($> 10$).
  A comparison with an unpublished MeerKat HI observation indicates a lower $\rm M_{HI}$-to-$\rm M_{mol}$ 
  ratio for the 6900 component ($\sim 4$).
In the HI map the 6600 component  fits well into a gap in the more extended 6000 component, albeit with a displacement of $\sim 5$~kpc.
\item[3]
  The three kinematic components in SQ-A have  very different SF activities. The 6600
  component has the highest SFR ($\rm 0.98\; M_\sun\; yr^{-1}$),
  dominated by the starburst (v$\sim 6680$~\kms in
$\rm H_\alpha$). The 6000 component has also a sizable SFR though about a
factor of 3 lower than that of 6600 component. The 6900
component has very little SF, with an SFR more than an order of
magnitude lower than that of the 6600 component.
\item[4]
  In order to probe at the cloud level the physical conditions that lead to
  the conspicuous difference in SFRs of different kinematic components,
  we carry out an analysis of CO clumps of sizes of $\sim 100$ -- 200~pc, utilizing
  the high resolution ALMA data (beam $= 0.38'' \times 0.23''$) and
  the clump-finding algorithm “Fellwalker” \citep{Berry2015}.
  The Larson relation between the effective size ($\rm R_e$) and the
  velocity dispersion ($\rm \sigma_v$) of these clumps are similar to that of
  the large Galactic GMCs.  No significant systematic
  difference is found among the different kinematic components
  in various physical parameters including
  $\rm R_e$, $\rm \sigma_v$, the
  virial parameter $\rm \alpha_{vir}$ and the external pressure $\rm P_e$.
\item[5]
  Correlation analyses of the clumps
  find no dependence of the SFR surface density $\rm \Sigma_{SFR}$
  and the depletion time of molecular gas $\rm t_{dep}$ on molecular gas surface density
  $\rm \Sigma_{mol}$ nor on the virial parameter $\rm \alpha_{vir}$, suggesting that
  the SFR enhancement (or suppression) is not due to any mechanism associated
  with the self-gravity. On the other hand, a marginally significant correlation is found between $\rm t_{dep}$ and $\rm \sigma_v$, suggesting that the gas motion may play a
  role in the SF activity of the clumps. Additional correlation analyses find 
  putatively significant correlation between  $\rm t_{dep}$ and the S parameter
  (a measure of projected angular momentum) and
  anti-correlation between  $\rm t_{dep}$ and the C parameter
  (a measure  of nonrotational-to-rotational energy ratio), suggesting
  that SFR may be suppressed by the centrifugal force in clumps with strong rotation
  and enhanced by compressive processes in clumps with high
  nonrotational-to-rotational energy ratios. No correlation is found between
  $\rm t_{dep}$ and the radial velocity gradient parameter $\rm G_v$, suggesting
  the share motion may not affect the SFR significantly.
\item[6]  
  In analog to the SF enhancement in the 30 Doradus plus southern ridge of the
  LMC \citep{Tsuge2024}, the following scenario for the triggering of the
  SQ-A starburst is supported by our results: A high-speed collision
  ($\rm v_{coll} \gsim 900$~\kms) is going on between the IGrM (pre-collision radial
  velocity $\sim 6900$~\kms) and a gas system associated with the intruder galaxy
  NGC~7319 (pre-collision radial velocity $\sim 6000$~\kms).
  Only a part of 6900 component and a part of 6000 component are
involved in the collision. During the collision the 6900 gas, which is
likely to be denser and more compact than the 6000 gas, is decelerated
and compressed, and this decelerated and compressed gas becomes the
6600 component in which the SQ-A starburst is
triggered. At the same time, the 6000 gas involved in the
collision is swept into the 6600 component and/or dispersed, leaving a
gap in the 6000 map at the place where the 6600 component crashes
through.  The $\sim 5$~kpc displacement shown in Fig.~\ref{fig:HI_shift}, which
is likely the projection of a 3-dimensional movement between the two components,
suggests an age of the collision of a
few $10^7$ years (assuming a transverse relative velocity of a few 100
\kms between the colliding systems), consistent with the estimated
time scale for the age of the encounter between the intruder and SQ
\citep{Sulentic2001}. According to simulations of \citet{Maeda2021, Maeda2024},
the high $\rm \Sigma_{SFR}$ and short $\rm t_{dep}$ of clumps in the 6600 component
(dominated by HI gas with a $\rm M_{HI}$-to-$\rm M_{mol}$ ratio $>10$)
are due to strong compression in the post-collision gas which triggers the formation
of a large amount of star-forming dense cores ($\rm n > 10^4 cm^{-3}$). 
\item[7]
  Different from the clumps of the 6600 component in SQ-A, the
  molecular clouds in the large scale shock region and in the 6900
  component in SQ-A show severely suppressed SF activity. These clouds
  are likely not involved in any cloud-cloud collision but in more complex collisions
  involving dense clouds and low density diffuse media.
  In this scenario, blast wave shocks (as shown
  by the radio and X-ray ridge) are triggered by the collision between
  diffuse media associated with IGrM and the intruder,
  respectively. As predicted by the model
  for interaction of blast wave shock and dense clouds
   \citep{Klein1994}, strong vorticity at the
  cloud-intercloud boundary and in the tail behind a dense cloud can be generated both
  by the initial passage of the blast wave shock and by the subsequent
  flow of post-shock inter-cloud diffuse gas passing-by the cloud. It is the
  vortices so generated that suppress the SF in the low SF activity clouds.
\end{description}

\begin{acknowledgments}
We thank K.~Rajpurohit for providing useful information from her MeerKAT observations before the publication. This work is sponsored (in part) by the Chinese Academy of Sciences (CAS) through a grant to the CAS South America Center for Astronomy. C.C. acknowledges NSFC grant No. 11803044 and 12173045. This work is supported by the China Manned Space Program with grant no. CMS-CSST-2025-A07. UL acknowledges support by the research grants PID2020-114414GB-I00 and PID2023-150178NB-I00, financed by MCIU/AEI/10.13039/501100011033  and from the Junta de Andaluc\'{i}a (Spain) grant  FQM108. P.A. and S.G. thank the Canadian Space Agency and the Natural Science and Engineering Council for support.
  
This paper makes use of the following ALMA data: ADS/JAO.ALMA\#2015.1.00241.S and ADS/JAO.ALMA\#2023.1.00177.S. ALMA is a partnership of ESO (representing its member states), NSF (USA) and NINS (Japan), together with NRC (Canada), MOST and ASIAA (Taiwan), and KASI (Republic of Korea), in cooperation with the Republic of Chile. The Joint ALMA Observatory is operated by ESO, AUI/NRAO and NAOJ. The National Radio Astronomy Observatory is a facility of the National Science Foundation operated under cooperative agreement by Associated Universities, Inc.

This work is based on observations made
with the NASA/ESA/CSA James Webb Space Telescope. The
data were obtained from the Mikulski Archive for Space
Telescopes at the Space Telescope Science Institute, which is
operated by the Association of Universities for Research in
Astronomy, Inc., under NASA contract NAS 5-03127 for JWST. 

Some of the data presented in this paper were obtained from the Mikulski Archive for Space Telescopes (MAST) at the Space Telescope Science Institute. The specific observations analyzed can be accessed via \dataset[https://doi.org/10.17909/9w9c-f533]{https://doi.org/10.17909/9w9c-f533}. STScI is operated by the Association of Universities for Research in Astronomy, Inc., under NASA contract NAS5–26555. Support to MAST for these data is provided by the NASA Office of Space Science via grant NAG5–7584 and by other grants and contracts.

(Some of) The data products presented herein were retrieved from the Dawn JWST Archive (DJA). DJA is an initiative of the Cosmic Dawn Center (DAWN), which is funded by the Danish National Research Foundation under grant DNRF140.

\end{acknowledgments}

\bibliographystyle{aasjournal}
\bibliography{sqa_revise_ax2.bbl}


\begin{appendix}
\section{Clumps in Field 4}\label{sec:A1}

We identified clumps in the CO(2-1) datacube of the "Field 4" observation of \citet{Appleton2023} using the FellWalker clump-finding algorithm via the {\sc PyCupid \footnote{\url{https://github.com/msanchezc/pycupid}}}, which is a Python wrapper for the Starlink CUPID package \citep{2007ASPC..376..425B, 2014ASPC..485..391C}. The detection parameters were configured as follows:
a detection threshold of 1-$\sigma$ above the background level, a minimum clump peak height of 2-$\sigma$ (minheight = 2), a separation threshold of 1-$\sigma$ between adjacent clumps (mindip = 1), a maximum step of 8 pixels to search for a higher neighboring pixel value (maxjump = 8), and a minimum clump size (this size is measured using the sum of all pixels in the clump without the flux weight) of twice the synthesized beam size in pixel (minpix = 2 Beam FWHM). These settings ensure detection of resolved, statistically significant structures while mitigating false positives in crowded molecular environments.
Clumps composited by two or more sub-clumps of spatially separated with gaps $\geq 5$~pixels (1~pixel$= 0.05''$) are excluded.
In total 36 clumps are identified in the field. Table~\ref{tab:clumps} lists various parameters of  these clumps. The parameters for individual clumps are defined as follows:

The position (i.e. R.A. \& Dec.) of each clump is specified by that
of its mass center whose x and y coordinates (in the image frame) are:
\begin{equation}
  \rm x =\Sigma_i f_i x_i / \Sigma_i f_i
\end{equation}
\begin{equation}
  \rm y =\Sigma_i f_i y_i / \Sigma_i f_i\; ,
\end{equation}
where $\rm x_i$, $y_i$ and $\rm f_i$ are the 
coordinates  and the
signal of the $\rm i_{th}$ pixel, respectively, and
the summation goes through all pixels in the clump. The parameter
$\rm r_{SQA}$ is the distance from a clump to the Field 4 center
(at $\rm RA = 338.995333$, $\rm Dec = +33.98074$)
which is pointed to the SQ-A starburst with a primary beam of  $22''$.
Two clumps (\#35 and \#36) have  $\rm r_{SQA} > 11''$ and therefore are outside the primary beam. They should be treated with caution.
The redshift velocity v and the velocity dispersion $\rm \sigma_v$ of a given clump are estimated as follows:
\begin{equation}
  \rm v = \Sigma_i f_i  v_i / \Sigma_i f_i \; ,
\end{equation}
\begin{equation}
  \rm \sigma_v = \sqrt{\Sigma_i f_i  (v_i - v)^2 / \Sigma_i f_i}.
\label{Eq:sigma_v}
\end{equation}
The effective radius of a clump is defined as
\begin{equation}
  \rm R_e = \sqrt{A_e/\pi}
\label{Eq:Re}
\end{equation}
where $\rm A_e$ is the signal weighted area of the clump in pc$^2$:
\begin{equation}
  \rm A_e = {a_{pix}  \Sigma_i f_i / f_{peak} \delta v \over \sqrt{2 \pi} \sigma_v}
  \label{Eq:Ae}
\end{equation}
where $\rm a_{pix}$ is the pixel area in pc$^2$, $\rm f_{peak}$ the peak
signal, and $\rm \delta v$ the velocity bin width.
Here we assume that in the velocity dimension signals follow a
Gaussian distribution characterized by $\rm \sigma = \sigma_v$.
The $\rm M_{mol}$ is estimated from the integrated flux $\rm f_{CO21}$ 
assuming a CO conversion factor
$\rm \alpha_{CO10} = 1.0\; M_\sun (K\; km\; s^{-1} pc^2)^{-1}$.
and a CO(2-1) to CO(1-0) luminosity ratio
$\rm r_{21} = L'_{CO(2-1)} /  L'_{CO (1-0)} = 0.34$ \citep{Emonts2025}.
The surface density $\rm \Sigma_{mol}= M_{mol} / A_e$.

We define a 2-D vector 
$\rm \vec{S} = S_x \vec{i} + S_y \vec{j}$  to represent the contribution of the radial velocity to the projected specific angular momentum (i.e. the angular momentum per unit mass
projected into the plane perpendicular to the line of sight). Here
$\rm S_x$ and $\rm S_y$ are the x- and y- components and $\rm \vec{i}$ and $\rm \vec{j}$ are unit vectors along the x and y axes, respectively (the cube is north-up, therefore the x axis directs toward west and y axis toward north).
The  absolute value of $\rm \vec{S}$, S, is a scalar parameter in the units of \kms~pc:
\begin{equation}
  \rm S = \sqrt{S_x^2+ S_y^2}
\label{Eq:S}
\end{equation}
where  
\begin{equation}
  \rm S_x = -d_{pix} {\Sigma_i f_i  (v_i - v) (y_i - y) \over  \Sigma_i f_i}
\end{equation}
\begin{equation}
  \rm S_y = d_{pix} {\Sigma_i f_i  (v_i - v) (x_i - x) \over  \Sigma_i f_i},
\end{equation}
 where $\rm d_{pix}$ is the pixel size in pc.
We also define a C parameter as follows:
\begin{equation}
  \rm C = (E - E_S)/ E_S
  \label{Eq:C}
\end{equation}
where E is a measure of the specific kinematic energy per unit mass:
\begin{equation}
  \rm E = {\sigma_v^2 \over 2}
\end{equation}
and $\rm E_s$ is a measure of the specific rotational energy:
\begin{equation}
  \rm E_S =  {S^2 \over 2J}  
\end{equation}
where J is a measure of the specific rotational inertia:
\begin{equation}
  \rm J = {d_{pix}^2 \Sigma_i f_i ((x_i - x)^2 + (y_i - y)^2) \over  \Sigma_i f_i}.
\end{equation}
By definition E$\rm \geq E_S$, therefore $\rm C \geq 0$.
Due to projection effects and beam smearing, E and $\rm E_S$ are not rigorous
representations of the total kinematic and rotational energies, respectively. Nevertheless C can be
used as an indicator of the nonrotational-to-rotational energy ratio in a clump.
When $\rm C < 1$
the kinematics in the clump is likely dominated by rotation and when $\rm C >> 1$
it is dominated by non-rotational motions.

In order to study the effects of shear motion, we define a 2-D
vector 
$\rm \vec{G_v} = G_{v,x} \vec{i} + G_{v,y} \vec{j}$  to represent the gradient of the radial
velocity. Its absolute value, $\rm G_{v}$, is a scalar parameter in the
units of \kms~$\rm pc^{-1}$:
\begin{equation}
  \rm G_v = \sqrt{G_{v,x}^2+ G_{v,y}^2}
  \label{Eq:Gv}
\end{equation}
where
\begin{equation}
  \rm G_{v,x} = {\Sigma_i w_i (v_i - v) /(x_i - x) \over d_{pix} \Sigma_i w_i}
\end{equation}
\begin{equation}
  \rm G_{v,y} = {\Sigma_i u_i (v_i - v) /(y_i - y) \over d_{pix} \Sigma_i u_i}.
\end{equation}
In order  to avoid run-away errors due to pixels with
coordinates too close to those of the mass center of the clump,
we choose the following weights (i.e. $\rm w_i$ and $\rm u_i$) in above summations:
\begin{eqnarray}
  \rm w_i  =  f_i \; (when (x_i-x)>0.5) \nonumber \\
  \rm      =  0   \; (when (x_i-x)\leq 0.5)
\end{eqnarray}
\begin{eqnarray}
  \rm u_i  = f_i \; (when (y_i-y)>0.5) \nonumber\\
  \rm      =  0   \; (when (y_i-y)\leq 0.5).
\end{eqnarray}

The external pressure $\rm P_e$ is calculated as follows \citep{Elmegreen1989}:
\begin{equation}
  \rm P_e = {3 \Gamma M_{mol} \sigma_v^2 \over 4\pi R_e^3}
    \label{Eq:Pe}
\end{equation}
where 
$\Gamma$ is the ratio between the density at the edge of a clump and the average
density. Following  \citet{Johnson2015}, we assume $\Gamma =0.5$. 

The virial parameter
$\rm \alpha_{vir}$ is defined by:
\begin{equation}
  \rm  \alpha_{vir} = {5\sigma_v^2R_e \over M_{mol}G}
    \label{Eq:avir}
\end{equation}
where $\rm G=4.3\times 10^{-3}$ is the gravitational constant in the units of  
pc~(\kms)$^2$~M$_\sun^{-1}$. 

The SFR surface density $\rm \Sigma_{SFR}$ is estimated using the JWST
15 $\mu m$ data.  Two operational issues make the estimation of
 $\rm \Sigma_{SFR}$ from integrated
15 $\mu m$ flux over the actual sky area of a clump unfavorable:
(1) many clumps are smaller than the
15 $\mu m$ beam; (2) on the 2-D image of the 15 $\mu m$ emission many
clumps are blinded with each other. Therefore we adopt a simpler
approach of measuring the
peak 15 $\mu m$ flux of each clump using a fixed aperture of $r=0.25''$ which
is approximately the resolution of the JWST 15 $\mu m$ image
(Section~\ref{sec:JWST}). The $\rm f_{15\mu m}$-to-SFR conversion is
calibrated using the total 15 $\mu m$ flux in the SQ-A region, which is
6.85~mJy, and the total $\rm  SFR = 1.42\; M_\sun\; yr^{-1}$
(sum of SFRs of the three components listed in Table~\ref{tab:cold_gas}).
This gives a $\rm f_{15\mu m}$-to-SFR
conversion factor of 0.21 $\rm M_\sun \; yr^{-1} mJy^{-1}$. 
Then $\rm \Sigma_{SFR}$ is estimated using the ratio between the SFR value and the aperture area:
\begin{equation}
 \rm \Sigma_{SFR}  = {SFR \over A_{aper}},
\end{equation}
where $\rm A_{aper}$ is the area of the aperture ($\rm r=0.25"$) in kpc$^2$.
Note that the SFR estimated from 15 $\mu m$ flux may be contaminated by
shocked gas emission of the $\rm [NIII]\lambda 15.56\mu m$ line
(Section~\ref{sec:JWST}).  However since the $\rm [NIII]\lambda
15.56\mu m$ line is rather weak in shocked gas (with $\rm
[NIII]/[NII]$ ratio about an order of magnitude lower than that in SF
regions, Appleton et al. in preparation), this contamination may be
insignificant.  Another uncertainty in using the 15 $\mu m$ emission as
an SFR indicator is due to possible destructions of small
grains which may result in underestimate of the SFR.
But, because we use the total SFR of SQ-A derived from extinction
corrected $\rm H_\alpha$ data \citep{DuartePuertas2021} instead of the one
measured from the 15 $\mu m$ luminosity, the adopted  $\rm f_{15\mu m}$-to-SFR
ratio should be unbiased on average. Nevertheless,
there can still be substantial clump-to-clump variations of this ratio. Given the
multiple uncertainties listed above, the error of the $\rm
\Sigma_{SFR}$ so estimated is on the order of a factor of 2.

The depletion time scale $\rm t_{dep}$ is another SF activity indicator, defined as follows:
\begin{equation}
 \rm t_{dep}  = {\Sigma_{mol} \over \Sigma_{SFR}}.
\end{equation}

\clearpage
\setcounter{table}{0}
\renewcommand{\thetable}{A\arabic{table}}
 
\begin{table*}
 \caption{Properties of individual CO clumps.}
\label{tab:clumps}
\resizebox{\textwidth}{!}{%
  \hspace*{-4.5cm}
  \begin{tabular}{lcccccccccccccccc}
    \noalign{\smallskip} \hline \noalign{\medskip}
    (1) & (2) & (3) & (4) & (5) & (6) & (7) & (8) & (9) & (10) & (11) & (12) & (13) & (14) &
    (15) & (16) & (17) \\
\\
    ID&  R.~A. & Dec.  & $r_{SQA}^a$& $v^b$   & $\sigma_v^c$ & $\rm R_e^d$
  & $\rm f_{CO21}^e$ &
  $\rm \log(M_{mol})^f$ &$\rm \alpha_{vir}^g$&$\rm \log(P_e/k_B)^h$&$\rm \log(\Sigma_{mol})^i$ &$\rm \log(S)^j$ &  $\rm \log(G_{v})^k$ &$\rm \log(C)^l$ &
$\rm \log(\Sigma_{SFR})^m$& $\rm \log(t_{dep})^n$ \\
\\
    & (J2000) & (J2000) & ($''$) & (\kms) & (\kms) & (pc) & (Jy~\kms) & (M$_\sun$) & &
    (K~$\rm cm^{-3}$) & (M$_\sun$~pc$^{-2}$) & (\kms pc) & (\kms pc$^{-1}$) & &
    (M$_\sun$~yr$^{-1}$kpc$^{-2}$) & (yr) 
\\
\noalign{\smallskip} \hline \noalign{\medskip}
  1& 338.99552&  +33.98058&  0.81& 6723&  5.13&  99.8& 0.096& 6.18&  2.0& 4.37& 1.68&  1.53& -2.05&  2.33& -0.17&  7.86\\
  2& 338.99535&  +33.98044&  1.11& 6717&  9.94&  76.6& 0.076& 6.07&  7.5& 5.18& 1.81&  1.59& -1.10&  2.68& -0.60&  8.41\\
  3& 338.99496&  +33.98070&  1.11& 6974&  7.46&  82.9& 0.067& 6.02&  5.1& 4.78& 1.69&  1.91& -1.55&  1.77& -1.25&  8.94\\
  4& 338.99503&  +33.98055&  1.13& 6750& 10.72& 104.6& 0.263& 6.61&  3.4& 5.38& 2.08&  2.82& -0.70&  0.54& -0.70&  8.78\\
  5& 338.99551&  +33.98044&  1.22& 6736&  4.22&  63.3& 0.023& 5.56&  3.6& 4.17& 1.46&  1.52& -1.36&  1.84& -0.35&  7.81\\
  6& 338.99507&  +33.98044&  1.37& 6747&  6.94&  75.7& 0.073& 6.06&  3.7& 4.88& 1.81&  1.61& -1.72&  2.30& -0.55&  8.36\\
  7& 338.99513&  +33.98035&  1.54& 6750&  5.47&  71.8& 0.043& 5.82&  3.7& 4.50& 1.62&  2.02& -1.34&  1.10& -0.84&  8.46\\
  8& 338.99551&  +33.98013&  2.30& 6909&  5.92&  59.9& 0.024& 5.58&  6.4& 4.56& 1.53&  1.62& -1.49&  1.76& -1.44&  8.97\\
  9& 338.99466&  +33.98030&  2.58& 6757&  8.07&  41.7& 0.014& 5.35& 14.0& 5.08& 1.62&  2.69& -0.66&  0.10& -1.46&  9.08\\
 10& 338.99507&  +33.98181&  3.92& 6926&  3.36&  70.0& 0.021& 5.51&  2.9& 3.79& 1.33&  1.68& -1.57&  1.44& -1.92&  9.25\\
 11& 338.99486&  +33.98186&  4.23& 6951&  7.80&  53.1& 0.035& 5.73&  7.0& 5.11& 1.79&  2.71& -0.85& -0.08& -1.83&  9.62\\
 12& 338.99599&  +33.98203&  5.02& 6957&  6.16&  64.0& 0.032& 5.69&  5.7& 4.62& 1.59&  1.56& -1.27&  1.92& -2.13&  9.71\\
 13& 338.99674&  +33.97996&  5.06& 6665&  5.02&  75.0& 0.047& 5.87&  3.0& 4.41& 1.62&  1.91& -1.33&  1.33& -1.51&  9.14\\
 14& 338.99471&  +33.98206&  5.08& 6949& 11.75&  80.2& 0.161& 6.40&  5.2& 5.59& 2.10&  2.19& -1.08&  1.62& -1.25&  9.35\\
 15& 338.99432&  +33.98217&  5.94& 6940& 17.12& 119.6& 0.629& 6.99&  4.2& 5.99& 2.34&  2.35& -1.57&  1.82& -1.08&  9.43\\
 16& 338.99641&  +33.98225&  6.29& 6908&  6.90&  56.1& 0.045& 5.85&  4.4& 5.05& 1.86&  2.10& -1.04&  0.96& -2.09&  9.94\\
 17& 338.99532&  +33.98261&  6.69& 6924&  7.56&  80.4& 0.096& 6.18&  3.6& 4.98& 1.87&  2.73& -0.97&  0.24& -1.82&  9.69\\
 18& 338.99653&  +33.98234&  6.75& 6935&  6.65&  70.0& 0.050& 5.89&  4.6& 4.77& 1.71&  2.16& -1.34&  1.30& -2.11&  9.82\\
 19& 338.99683&  +33.98217&  6.79& 6919&  8.41&  87.7& 0.126& 6.29&  3.7& 5.08& 1.92&  2.62& -0.95&  0.64& -2.26& 10.18\\
 20& 338.99553&  +33.98278&  7.35& 6940&  8.93& 102.3& 0.218& 6.53&  2.8& 5.17& 2.02&  2.47& -1.14&  1.17& -1.33&  9.35\\
 21& 338.99546&  +33.98285&  7.58& 6943&  5.17&  66.1& 0.041& 5.80&  3.2& 4.54& 1.67&  1.96& -1.62&  1.12& -0.91&  8.58\\
 22& 338.99286&  +33.98020&  7.65& 6033&  4.63&  98.9& 0.074& 6.06&  2.1& 4.18& 1.58&  2.01& -1.54&  1.21& -1.25&  8.83\\
 23& 338.99572&  +33.98288&  7.77& 6934& 12.16&  45.1& 0.027& 5.63& 18.3& 5.60& 1.83&  2.73& -0.84&  0.28& -1.80&  9.63\\
 24& 338.99469&  +33.98287&  7.87& 6947&  6.22&  86.1& 0.101& 6.20&  2.5& 4.75& 1.83&  2.33& -1.34&  0.84& -1.87&  9.70\\
 25& 338.99267&  +33.98020&  8.20& 6011& 10.87&  56.3& 0.071& 6.05&  7.0& 5.63& 2.05&  2.49& -1.68&  1.05& -1.41&  9.46\\
 26& 338.99569&  +33.98305&  8.37& 6894& 13.08&  62.4& 0.087& 6.13&  9.1& 5.75& 2.05&  2.68& -0.43&  0.60& -1.78&  9.83\\
 27& 338.99593&  +33.98308&  8.60& 6927&  7.10&  77.9& 0.119& 6.27&  2.5& 5.06& 1.99&  2.22& -1.70&  1.06& -1.99&  9.98\\
 28& 338.99584&  +33.98311&  8.63& 6930&  3.77&  58.1& 0.027& 5.62&  2.3& 4.25& 1.60&  0.95& -1.40&  3.04& -1.96&  9.55\\
 29& 338.99253&  +33.98010&  8.68& 6008&  6.31&  47.6& 0.030& 5.67&  4.7& 5.01& 1.82&  2.40& -0.77&  0.31& -1.27&  9.09\\
 30& 338.99605&  +33.98316&  8.94& 6868&  3.09&  69.0& 0.023& 5.55&  2.2& 3.78& 1.38&  1.81& -2.15&  1.02& -2.02&  9.40\\
 31& 338.99232&  +33.98011&  9.28& 6011&  2.69&  73.4& 0.026& 5.60&  1.5& 3.63& 1.37&  1.89& -1.47&  0.72& -0.98&  8.35\\
 32& 338.99617&  +33.98324&  9.30& 6871&  6.35&  48.1& 0.023& 5.55&  6.4& 4.88& 1.69&  2.55& -0.87& -0.12& -2.01&  9.70\\
 33& 338.99590&  +33.98333&  9.45& 6903& 12.01&  72.0& 0.159& 6.39&  4.9& 5.75& 2.18&  2.84& -0.93&  0.30& -1.98& 10.16\\
 34& 338.99596&  +33.98334&  9.51& 6882&  7.03&  69.3& 0.064& 6.00&  4.0& 4.94& 1.82&  2.25& -1.92&  1.04& -2.01&  9.84\\
 35& 338.99166&  +33.98205& 11.91& 6922&  7.04&  46.0& 0.024& 5.57&  7.1& 5.04& 1.75&  2.00& -1.20&  1.30& -1.74&  9.49\\
 36& 338.99092&  +33.97975& 13.64& 6006&  4.29&  66.3& 0.073& 6.06&  1.2& 4.63& 1.92&  1.89& -1.11&  1.24& -1.30&  9.21\\
\\
\noalign{\smallskip} 
\hline \noalign{\medskip}
\end{tabular}}

{\small
\noindent{\bf Notes:} 
$^a$ Distance from the center of the SQ-A starburst ($\rm RA = 338.995333$, $\rm Dec = +33.98074$).
$^b$ Redshift velocity.
$^c$ Velocity dispersion.
$^d$ Effective radius.
$^e$ Integrated CO(2-1) flux, with a $\sim 15\%$ error dominated by the calibration uncertainty.
$^f$ Logarithm of $\rm M_{mol}$, with an error on the order
of a factor of $\sim 2$ dominated by the uncertainty of the conversion
factor $\rm \alpha_{CO10}$.
$^g$ Virial parameter.
$^h$ Logarithm of the external pressure divided by the Boltzmann constant. 
$^i$ Logarithm of the molecular gas surface density.
$^j$ Logarithm of the S parameter  (a measure of the projected specific angular momentum).
$^k$ Logarithm of the gradient of radial velocity.
$^l$ Logarithm of the C parameter  (a measure of the nonroational-to-rotational energy ratio).
$^m$ Logarithm of the SFR surface density.
$^n$ Logarithm of the depletion time scale of the molecular gas ($\rm t_{dep} = \Sigma_{mol} / \Sigma_{SFR}$).
}
\end{table*}

\end{appendix}
\end{document}